\documentclass[aps,prb,twocolumn,notitlepage,showpacs,superscriptaddress,10pt]{revtex4-1}%
\usepackage{graphicx}
\usepackage{amsmath}
\usepackage{bm}
\usepackage{amssymb}
\usepackage{color,soul}
\usepackage{amsfonts}%
\usepackage[caption=false]{subfig}
\usepackage{comment}
\usepackage{color,soul}
\usepackage[dvipsnames]{xcolor}
\usepackage[capitalize]{cleveref}
\usepackage{float}
\usepackage{tabularx}
\usepackage{array}   
\newcolumntype{L}{>{$}l<{$}} 
\newcolumntype{R}{>{$}r<{$}} 
\newcolumntype{C}{>{$}c<{$}} 
\usepackage{comment}
\usepackage[capitalize]{cleveref}

\definecolor{brown}{rgb}{0.7,0,0}

\date{\today}

\newcommand{\frankfurt}{Institut f\"ur Theoretische Physik, Goethe-Universit\"at Frankfurt,
Max-von-Laue-Strasse 1, 60438 Frankfurt am Main, Germany}
\newcommand{\paris}{CPHT, CNRS, Ecole Polytechnique, Institut Polytechnique de Paris, Route de Saclay, 91128 Palaiseau, France,\\
F-91128 Palaiseau, France Coll\'ege de France, 11 place Marcelin Berthelot, 75005 Paris,\\
France European Theoretical Spectroscopy Facility, 91128 Palaiseau, France, Europe}
\newcommand{\salem}{Department of Physics and Center for Functional Materials, Wake Forest University, Winston-Salem, North Carolina 27109, United States}
\newcommand{\cnrspin}{Consiglio Nazionale delle Ricerche CNR-SPIN, c/o Universit{\'a} degli Studi 'G. D’Annunzio’, 66100, Chieti, Italy}
\newcommand{\uliege}{NanoMat/Q-mat/CESAM,Universit\'e de Li\`ege, B-4000 Li\`ege, Belgium}
\newcommand{\cnrspinRome}{Consiglio Nazionale delle Ricerche CNR-SPIN, Area della Ricerca di Tor Vergata,Via del Fosso del Cavaliere 100, I-00133 Rome, Italy}
\newcommand{\sanken}{Institute of Scientific and Industrial Research ISIR-SANKEN, Osaka University, 8-1 Mihogaoka, Ibaraki, Osaka, 567-0047, Japan}
\newcommand{\osaka}{Department of Precision Engineering, Graduate School of Engineering, Osaka University, 2-1 Yamadaoka, Suita, Osaka 565-0871, Japan}

\begin{document}

\title{Microscopic origin of magnetism in monolayer $3d$ transition metal dihalides}

\author{Kira Riedl}
\thanks{These two authors contributed equally}
\affiliation{\frankfurt}
\author{Danila Amoroso}
\thanks{These two authors contributed equally}
\affiliation{\cnrspin}
\affiliation{\uliege}
\author{Steffen Backes}
\affiliation{\paris}
\author{Aleksandar Razpopov}
\affiliation{\frankfurt}
\author{\\Thi Phuong Thao Nguyen }
\affiliation{\sanken}
\affiliation{\osaka}
\author{Kunihiko Yamauchi}
\affiliation{\sanken}
\affiliation{\osaka}
\author{Paolo Barone}
\affiliation{\cnrspinRome}
\author{Stephen M. Winter}
\affiliation{\salem}
\author{Silvia Picozzi}
\affiliation{\cnrspin}
\author{Roser Valent{\'\i}}
\affiliation{\frankfurt}

\begin{abstract}
Motivated by the recent wealth of exotic magnetic phases emerging in two-dimensional frustrated lattices, we investigate the origin of possible magnetism in the monolayer family of triangular lattice materials $MX_2$ ($M$=\{V, Mn, Ni\}, $X$=\{Cl, Br, I\}). We first show that consideration of general properties such as filling and hybridization enables to formulate trends for the most relevant magnetic interaction parameters.  In particular, we observe that the effects of spin-orbit coupling (SOC) can be effectively tuned through the ligand elements as the considered 3$d$ transition metal ions do not strongly contribute to the anisotropic component of the inter-site exchange interaction. Consequently, we find that the corresponding SOC matrix-elements differ significantly from the atomic limit. In a next step and by using two {\it ab initio}-based complementary approaches,
 we extract realistic effective spin models and find that in the case of heavy ligand elements, SOC effects manifest in anisotropic exchange and single-ion anisotropy only for specific fillings.
\end{abstract}

\maketitle

\section{Introduction}

Transition-metal-based materials with magnetically frustrated lattices
have been at the center of intensive research for several decades\cite{ramirez1994strongly,balents2010spin,lacroix2011introduction,starykh2015unusual,liu2019spintronic,vasiliev2019low} due to the presence of fascinating phases ranging from unconventional ordered states to spin liquids. In recent years, two-dimensional (2D) van der Waals magnets have emerged as a new platform for exotic magnetism in reduced dimensions. One of the most prominent examples is the honeycomb spin-1/2 $\alpha$-RuCl$_3$ which has dominant frustrating Kitaev interactions\cite{winter2016challenges,winter2017breakdown,kim2016crystal,yadav2016kitaev,hou2017unveiling,eichstaedt2019deriving,laurell2020dynamical}
 as a result of an interplay of crystal field splitting, Coulomb repulsion and spin-orbit coupling of Ru 4$d$ electrons. 
 Anisotropic exchange interactions are also being 
 discussed in the context of 2D van der Waals magnets with 3$d$ transition metals, such as the spin-3/2 CrI$_3$, 
 where SOC effects arise from the ligand iodine mediating the exchange; nevertheless,
 the underlying physics and strength of such interactions still remain rather controversial\cite{lado2017origin,xu2018interplay,besbes2019microscopic,lee2020fundamental,kartsev2020biquadratic,kvashnin2020relativistic,stavropoulos2021magnetic,edstrom2021curved}.
 
  \begin{figure}
    \centering
    \includegraphics[width=\linewidth]{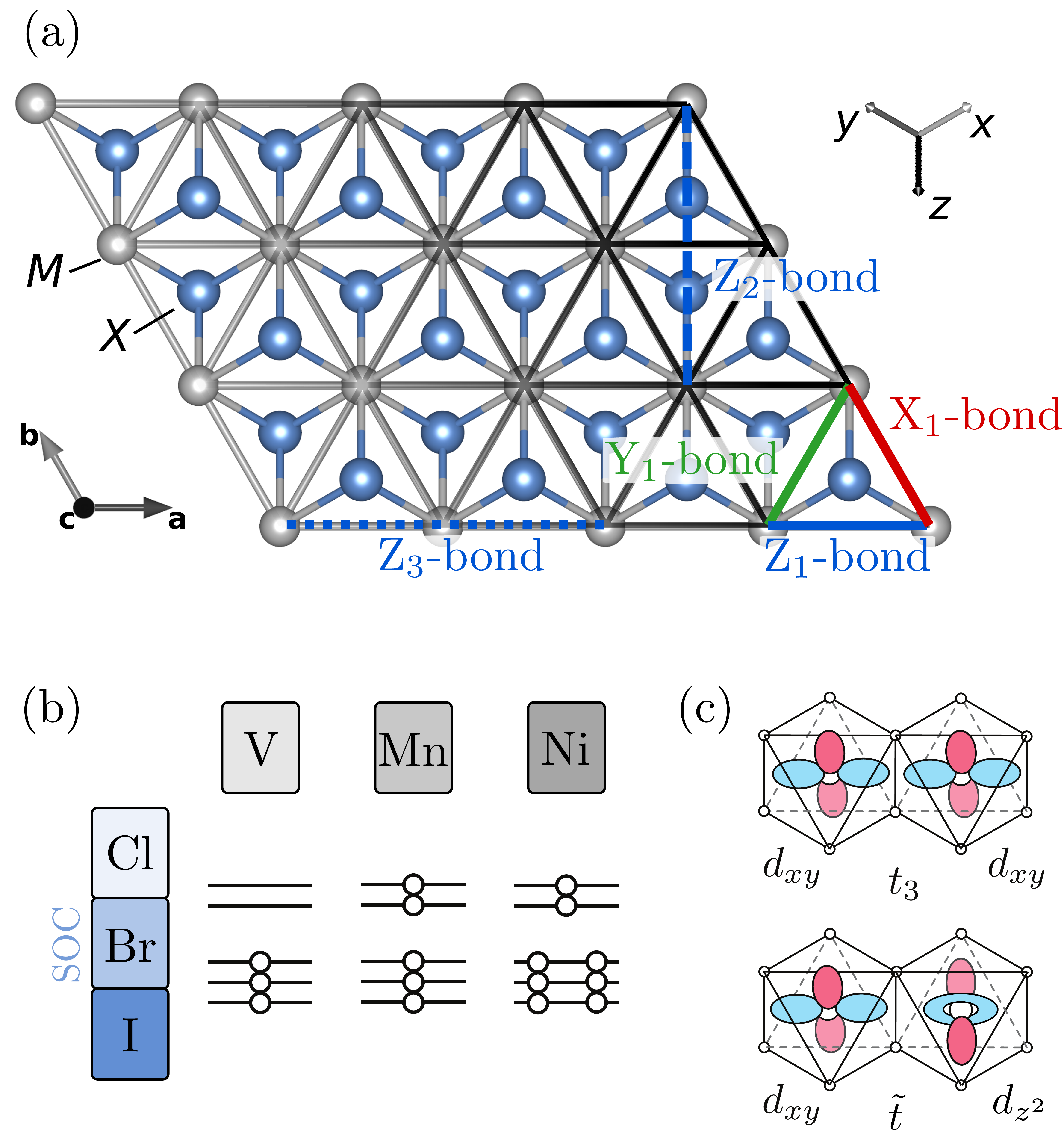}
    \caption{(a) Common structure of the triangular $MX_2$ materials, for which monolayer structures are investigated in this work. Relevant bonds, the crystallographic $(abc)$ coordinate system and the cubic $(xyz)$ coordinate system are defined.
    (b) Filling for V$X_2$, Mn$X_2$ and Ni$X_2$ with $X=\{\text{Cl,\,Br,\,I}\}$. A darker colour illustrates higher filling (grey) or stronger SOC (blue). (c) Dominant hoppings $t_3$ and $\tilde{t}$ on a nearest neighbor Z$_1$-bond.
    }
    \label{fig:structure}
\end{figure}

Motivated by  the significance of understanding the
origin of mechanisms behind the different magnetic interactions in 2D van der Waals (vdW) magnets, we study here the case of magnetic exchange interactions of selected monolayer $3d$ transition metal dihalides $MX_2$ ($M$=\{V, Mn, Ni\}, $X$=\{Cl, Br, I\}), where the cations $M$ are in the octahedral environment of the
 ligand $X$ anions, and are arranged in a triangular lattice, as depicted in \cref{fig:structure}(a).
 These materials exhibit a variety of interesting properties. For example, type-II multiferroicity was reported in bulk NiBr$_2$\cite{tokunaga2011multiferroicity}, NiI$_2$\cite{kurumaji2013magnetoelectric} and MnI$_2$\cite{kurumaji2011magnetic,li2020magnetic}, 
 where the onset of helimagnetic phases breaks inversion symmetry leading to spin-induced ferroelectricity. Such a multiferroic phase has been recently reported to survive down to the monolayer limit of NiI$_2$ \cite{song2022experimental,Fumega_2022}, while monolayer NiBr$_2$ and NiCl$_2$ have been proposed as half-excitonic insulators\cite{jiang2019half}.
 The electronic structure of the $3d$ $MX_2$ compounds has been recently analyzed \cite{botana2019electronic,stavropoulos2019microscopic}; however 
 the magnetic exchange couplings and SOC-driven anisotropic contributions have not been fully addressed and require deeper investigation.

 The purpose of the present work is twofold: (i) to uncover the microscopic mechanism behind
the magnetic couplings in the 2D dihalides with $3d$ transition metals, and (ii) to benchmark  two complementary \textit{ab-initio} approaches for the estimation of magnetic couplings, the \mbox{``projED''}\cite{riedl2019abinitio} and ``4-states''\cite{xiang2013magnetic,li2021spin} methods.
The selected materials allow us to address the underlying processes behind the magnetic couplings
by studying the influence of different electron filling through the metal elements $M$, and the effective spin-orbit coupling (SOC) through the ligand elements $X$, as schematically
shown in \cref{fig:structure}(b).  
The magnetic properties of the respective materials can be then generally described in terms of effective (anisotropic) spin Hamiltonians with $S=$\{3/2, 5/2, 1\}, dependent on the filling \{$d^3$, $d^5$, $d^8$\}. 

The paper is organized as follows; in \cref{sec:model,sec:trends} we discuss the spin model and general trends for the magnetic exchange couplings in $MX_2$ considering crystal-field splittings and the Goodenough-Kanamori-Anderson rules\cite{Anderson1950,Kanamori1958}. In \cref{sec:MX2_results} we  present
our results on the \emph{ab-initio} estimated magnetic interactions and discuss the resulting
magnetic properties for the dihalide family.  In \cref{sec:MX2_results} we
present our conclusions.

\section{Spin Model}
\label{sec:model}

The most general spin Hamiltonian model, including single-ion anisotropy $\mathbb{A}_l$ and bilinear exchange tensors $\mathbb{J}_{lm}$ (with sites $l$, $m$), 
is given by:
\begin{align}
\mathcal{H}=\sum_l \mathbf{S}_l \cdot \mathbb{A}_{l} \cdot \mathbf{S}_l + \sum_{<lm>} \mathbf{S}_l \cdot \mathbb{J}_{lm} \cdot \mathbf{S}_m.
\label{hamiltonian_eqn}
\end{align}
For convenience, we express the single-ion anisotropy (SIA) in the crystallographic coordinate system ($ab^\ast c$), where $b^\ast$ is perpendicular to the $ac$ crystal axes, $\mathcal{H}_{\rm SIA}=A_c \sum_l \, (S_l^{c})^2$ consistently with the symmetry of MX$_2$. This allows us to easily identify the triangular layer as an easy or hard plane. On the other hand, the bond-dependent  bilinear exchange 
parameters are most conveniently described in the cubic coordinate system ($xyz$), which consists of orthogonalized axes oriented approximately along $M$-$X$ bonds, as illustrated in \cref{fig:structure}(a). Note that the $M$-$X$ bonds are not perfectly orthogonal due to trigonal distortion effects. 

The $P\bar{3}m1$ (space group 164) centrosymmetric monolayer structures 
have four independent parameters. Here, we adopt a bond-dependent parametrization corresponding to the extended Heisenberg-Kitaev model \cite{kitaev2006anyons,winter2017models}.
In this framework, each bond is labelled after the cubic axis perpendicular to it. For a nearest-neighbor Z$_1$-bond, perpendicular to the $z$ axis, (\cref{fig:structure}(a)), the  exchange couplings are then conventionally parametrized as follows: 
\begin{align}
\mathbb{J}_{lm} = \left(\begin{array}{ccc} J & \Gamma & \Gamma^\prime \\ \Gamma & J & \Gamma^\prime \\ \Gamma^\prime & \Gamma^\prime & J+K \end{array} \right).
\label{eq:interaction_tensor}
\end{align}
with the (nearest neighbor) bond-isotropic Heisenberg exchange $J$, the anisotropic Kitaev exchange $K$, and  the off-diagonal symmetric exchange terms $\Gamma$ and $\Gamma^\prime$.
This is the most symmetric general expression for the exchange tensor given the crystal symmetries that enforce four independent parameters.
 The bilinear exchange matrices for the X- and Y-bond are related to \cref{eq:interaction_tensor} by $C_3$ rotation about the out-of-plane axis. A correspondence to a parametrization oriented along the crystallographic axes coordinate system $(ab^\ast c)$ is given in \cref{sec:coordinates}.

\section{General Considerations}
\label{sec:trends}

We start by identifying general traits in hybridization patterns and fillings in $3d$-$MX_2$ triangular
lattice compounds, with edge-sharing halogen ligand octahedra. 
As we discuss below, this will allow us to formulate trends for the most relevant magnetic interactions.

\subsection{Metal-ligand hybridization}
\label{sec:metalligand}

The hybridization between the metal $d$- and ligand $p$-orbitals is generally expected to increase as the electronegativity difference between the two elements decreases. This affects related electronic properties, such as the local Coulomb repulsion $U_{\rm avg}$, the strength of different hopping processes and the materials' SOC matrix elements. These factors all together ultimately determine the magnetic couplings.

\begin{figure}
    \centering
    \includegraphics[width=\linewidth]{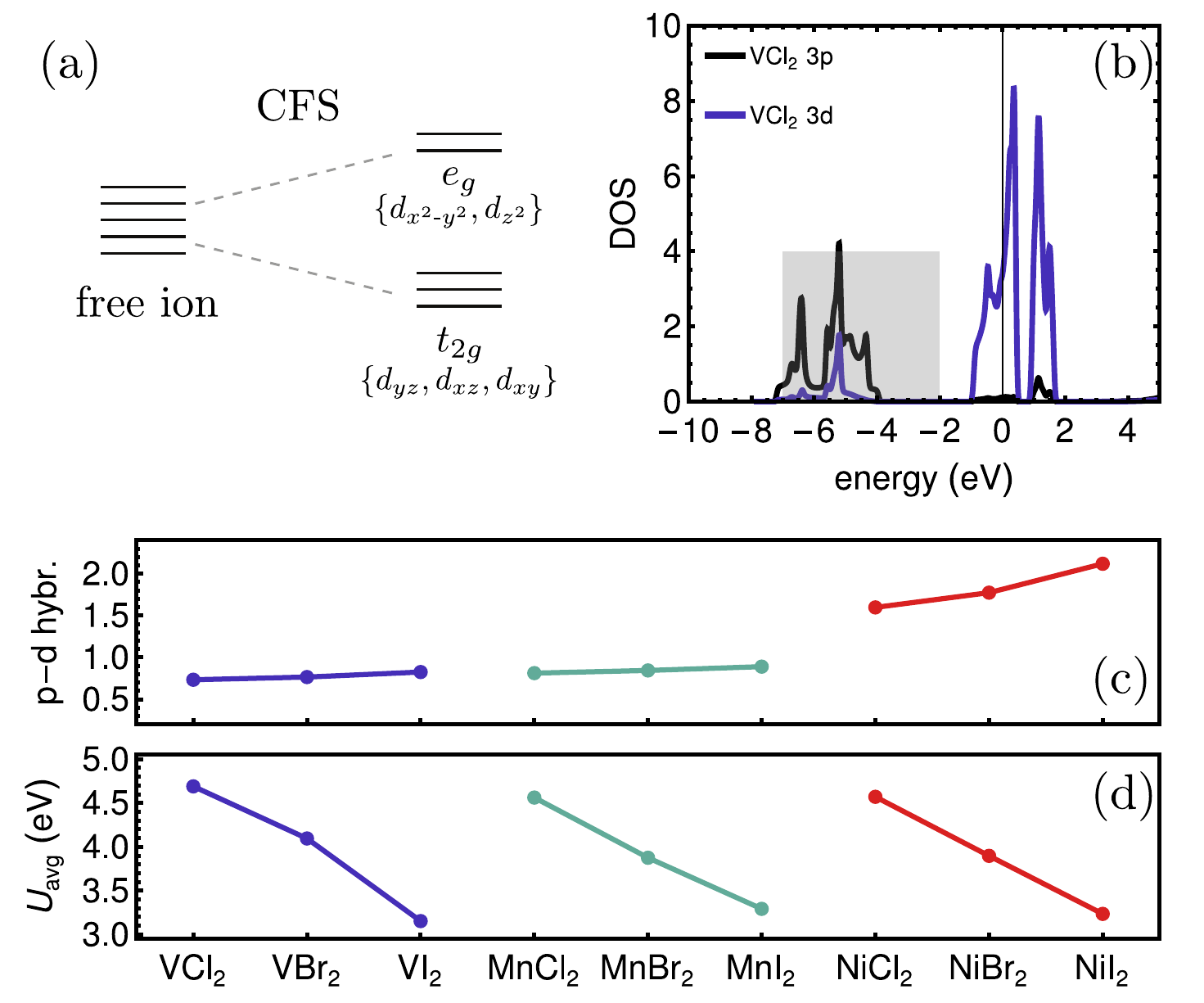}
    \caption{(a)
    Cubic crystal-field splitting (CFS) of $d$ orbitals into $t_{2g}$ and $e_g$ levels, (b) \textit{ab-initio} (GGA) density of states (DOS) of VCl$_2$, (c) $p$-$d$ hybridization, i.e. integral of $3d$ DOS in the energy region dominated by $p$-$d$ bonding (illustrated by the grey box in (b)), and (d) $U_{\rm avg}$ calculated with cRPA for each $MX_2$ material. }
    \label{fig:uvals}
\end{figure}

We quantify the hybridization in $MX_2$ via non spin-polarized density functional theory (DFT) calculations, through the generalized gradient approximation (GGA)~\cite{pbe} orbital-resolved density of states (DOS).
In particular, the energy window dominated by $p$-$d$ bonding orbitals contains also a finite $d$ character that scales with the degree of hybridization. This is illustrated for the example case of VCl$_2$ in \cref{fig:uvals}(b), where a finite $d$ character (in blue) is present within the relevant energy window (marked by the grey box). 
The integral of the metal $3d$ DOS in the respective $p$-$d$ bonding dominated energy window is shown for each material in \cref{fig:uvals}(c). 
As expected, the hybridization increases with smaller electronegativity difference, i.e. for each of the considered metal elements it increases with ligand atomic number. For the Ni$X_2$ systems, an especially strong hybridization is observed due to the larger electronegativity of nickel.

To estimate trends in the effective Coulomb interaction $U_{\rm avg}$, we employed the constrained random-phase approximation (cRPA)~\cite{Aryasetiawan2004,Aryasetiawan2006}, detailed in \cref{sec:methods}.
The results for each material are given in \cref{tab:cRPA}, and $U_{\rm avg}$ is illustrated in \cref{fig:uvals}(d). We observe a decrease of $U_{\rm avg}$ as a function of ligand atomic number, which can be related to the increasing metal-ligand hybridization (see \cref{fig:uvals}(c)). Accordingly, we find the smallest effective interaction parameters for the strongest hybridizing, and hence most delocalized, orbitals. 
We have taken the cRPA values of the effective Coulomb repulsion $U_{\rm avg}$ and Hund's coupling $J_{\rm avg}$ as input modeling parameters to estimate the magnetic interaction parameters by means of the two \textit{ab-initio} methods (\cref{sec:MX2_results}). 

\begin{table}[b]
    \centering\def\arraystretch{1.1}
    \begin{ruledtabular}
    \begin{tabular}{C|CCC|CCC|CCC}
    & \multicolumn{3}{C|}{\text{V}X_2} & \multicolumn{3}{C|}{\text{Mn}X_2}& \multicolumn{3}{C}{\text{Ni}X_2}  \\
    X & \text{Cl} & \text{Br} & \text{I} & \text{Cl} & \text{Br} & \text{I} & \text{Cl} & \text{Br} & \text{I} \\
    \hline
    U_{\rm avg}  &4.69 &4.10 &3.15 &4.56 &3.88 &3.29 &4.57 &3.90 &3.24 \\
    J_{\rm avg}  &0.65 &0.63 &0.57 &0.79 &0.75 &0.71 &0.84 &0.79 &0.68
    \end{tabular}
    \end{ruledtabular}
    \caption{Constrained RPA results of averaged Hubbard repulsion $U_{\rm avg}$ (in eV) and averaged Hund's coupling $J_{\rm avg}$ (in eV) for the monolayer structures of each investigated material.}
    \label{tab:cRPA}
\end{table}

\subsection{Nearest Neighbor Isotropic Exchange Interactions}
\label{sec:nearest_exchange}

In general, there are various exchange processes that are relevant to the magnetic couplings. The trends for different ligands and filling can be rationalized by considering effective $d$-$d$ hopping integrals estimated via Wannier projection\cite{pavarini2012,eschrig2009tight} onto a $d$-only basis. Precise hopping parameters estimated using non-relativistic DFT calculations with the full potential local orbital (FPLO)~\cite{fplo} 
basis as well as complete expressions for symmetry-allowed hopping matrices are given in \cref{sec:hopping}. An alternative analysis, based on spin-polarized Wannier projection is detailed in \cref{sec:spinWan}.

Over the entire series of materials, we find that the dominant nearest neighbor hoppings are $t_3$ and $\tilde{t}$; 
on the Z$_1$-bond, for example, these correspond to $t_{(xy;xy)}$ and $t_{(xy,z^2)}$, respectively, illustrated in \cref{fig:structure}(c).
For the ideal 90$^\circ$ $M$-$X$-$M$ bond geometry, $t_3$ is mainly the result of direct overlap, while $\tilde{t}$ has contributions from both direct hopping and ligand hybridization (with the latter dominating). 
Consequently, $\tilde{t}$ becomes increasingly important as the $p$-$d$ hybridization increases.

The consequences of the dominant $d$-$d$ hopping parameters $t_3$ and $\tilde{t}$ on the magnetic interactions can be understood in terms of filling of the $t_{2g}$ or $e_g$ orbitals (see \cref{fig:uvals}(a)) according to the Goodenough-Kanamori-Anderson (GKA) rules\cite{Anderson1950,Kanamori1958}. In particular, hopping between half-filled orbitals is associated with antiferromagnetic exchange, while hopping from a half-filled to a full or empty orbital is associated with ferromagnetic exchange.

For the $d^8$ materials (i.e. Ni$X_2$), we generally expect the nearest neighbor couplings to be ferromagnetic: the $t_{2g}$ levels are filled; hence, the hopping processes involving $t_3$ do not contribute to exchange in lowest order. 
Therefore, $J_1$ arises mainly from $\tilde{t}$ processes, which connect a half-filled $e_g$ orbital to a filled $t_{2g}$ orbital, and leads to a ferromagnetic exchange. This effect is enhanced in systems with heavier ligands, as the ligand-assisted $\tilde{t}$ is strengthened by increased $p$-$d$ hybridization.

For the $d^3$ materials (i.e. V$X_2$), there is a competition between ferromagnetic (FM) and antiferromagnetic (AFM) contributions to the nearest neighbor exchange.
In addition to the ferromagnetic exchange arising from the $\tilde{t}$ process connecting an empty $e_g$ with a half-filled $t_{2g}$ orbital, an antiferromagnetic contribution arises from the hopping between the two half-filled $t_{2g}$ levels via $t_3$. While both mechanisms are relevant, the ferromagnetic contributions are typically weaker than antiferromagnetic contributions, so that $J_1 > 0$ (in part, because $|t_3| > \tilde{t}$). 
For heavier ligands, the increasing $\tilde{t}$ should enhance the ferromagnetic contribution primarily, resulting in a decreased magnitude of $J_1$.

Finally, for the high spin $d^5$ case (i.e. Mn$X_2$), the overall magnitude of the couplings is generally expected to be weak. While both $\tilde{t}$ and $t_3$ hopping processes contribute with antiferromagnetic contributions due to half-filled $t_{2g}$ and $e_g$ levels, their effects are reduced by two main factors. First, 
the energy cost for transfer of an electron between metal sites is the largest for the case of half filling\cite{georges2013strong}. This can be understood by considering the effects of Hund's coupling; for a process $(\uparrow\uparrow\uparrow\uparrow\uparrow,\downarrow\downarrow\downarrow\downarrow\downarrow)\to (\uparrow\uparrow\uparrow\uparrow,\downarrow\downarrow\downarrow\downarrow\downarrow\uparrow)$, the total Coulomb repulsion experienced by the electron at its parent site is mitigated by the spin alignment through Hund's coupling. After hopping, the electron's spin is necessarily anti-aligned with all other electrons, so that the full Coulomb repulsion is felt. For this reason the cost for electron transfer is large $\sim U + 4J$. This suppresses the antiferromagnetic exchange sufficiently that ferromagnetic exchange processes not captured in the $d$-only picture are competitive, resulting in an overall suppression of the nearest neighbor interactions. Full discussion of this situation is provided in the following sections.

\subsection{Anisotropic Interactions}

In this work, we consider $d^3$, high-spin $d^5$, and $d^8$ filling because the ground states possess no orbital degeneracy, and the orbital angular momentum is quenched at zeroth order. In addition, the atomic spin-orbit coupling for third row metals is relatively weak. For this reason, magnetic anisotropy associated with the metal alone is mostly negligible, allowing the effects of introducing heavy ligands to be investigated in detail. We discuss these effects again in terms of the effective $d$-only model, where the atomic SOC of the ligands ($\mathcal{H}_{\rm SOC} = \xi \sum_l \mathbf{L}_l \cdot \mathbf{S}_l$) is downfolded into effective $d$-$d$ hopping and on-site terms. Such terms can be estimated via Wannier projection techniques applied to fully relativistic DFT calculations, as shown in the following sections. In general, such single-particle contributions then take the form:
\begin{align}
\mathcal{H}_{\rm hop}=\sum_{lm} \sum_{\alpha \beta} \underline{c}_{l\alpha}^{\rm T}\ \{ t_{\alpha \beta}^{lm}\ \mathbb{I} + \frac{i}{2} (\vec{\lambda}_{\alpha \beta}^{lm} \cdot \vec{\sigma})\}\ \underline{c}_{m \beta}
\end{align}
in terms of the Pauli matrices $\sigma$ and electron annihilation
operators on site $l$ and orbital $\alpha$, $\underline{c}_{l \alpha} = \begin{pmatrix} c_{l \alpha \uparrow} & c_{l \alpha \downarrow} \end{pmatrix}$
and its transpose $\underline{c}_{l \alpha}^{\rm T}$. Here, the $t_{\alpha\beta}^{lm}$ represent the regular spin-independent hopping and crystal field terms between orbital $\alpha$ at metal site $l$ and $\beta$ at site $m$. The vector $\vec{\lambda}_{\alpha\beta}^{lm}$ then parameterizes the complex spin-dependent terms resulting from spin-orbit coupling. For example, the contribution from the atomic SOC at the metal site M corresponds to $\vec{\lambda}_{\alpha\beta}^{ll} = -i \ \xi_\text{M} \langle \alpha | \mathbf{L} |\beta\rangle$. With inclusion of heavy ligands with large SOC constants $\xi$, there are two main effects. 

The first effect is to induce spin-dependent crystal field terms as a result of $p$-$d$ hybridization that mimic the atomic SOC. For example, the metal $d_{xy}$ orbital may hybridize with the ligand $p_y$ orbital, while the $d_{xz}$ hybridizes with the ligand $p_z$ orbital. The effects of SOC at the ligand is then to make an effective matrix element between the $d$-orbitals via the sequence $d_{xy} \ \overrightarrow{_{\text{hop}}}\ p_y \ \overrightarrow{_{L_xS_x}} \ p_z \ \overrightarrow{_{\text{hop}}} \ d_{xz}$. When downfolded into the $d$-only picture, this mimics the effects of $L_xS_x$, with an effective SOC constant that scales with the atomic constant of the ligand and the degree of metal-ligand hybridization. However, as discussed further in \cref{sec:hopping}, the matrix elements of the induced SOC are not restricted to take the atomic form. They are the primary source of single-ion anisotropy (SIA), but the effects are difficult to anticipate {\it a priori}. We can anticipate only that the SIA should generally grow for heavier ligands. Experimentally, $3d^3$ and $3d^5$ materials tend to have weak single-ion anisotropy, while larger variations are seen for $3d^8$ materials\cite{abragam2012electron}. 

\begin{figure}
    \centering
    \includegraphics[width=\linewidth]{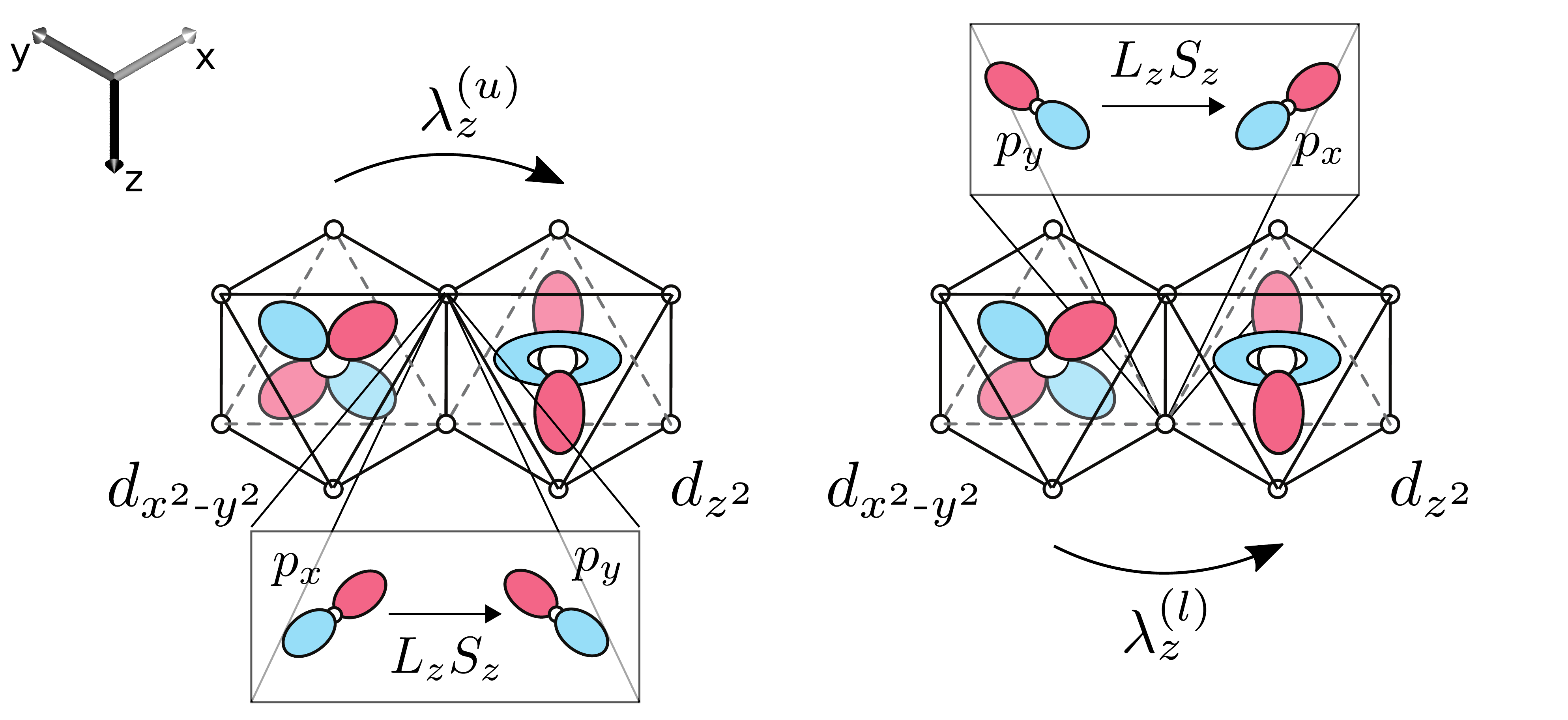}
    \caption{Illustration of predominant nearest-neighbor spin-dependent hopping $\lambda_z$, arising from spin-orbit coupling effects on a Z$_1$-bond (defined in \cref{fig:SOC_hopping}). It results from the hopping along the upper ($\lambda_z^{(u)}$) and lower ($\lambda_z^{(l)}$) paths: $\lambda_z = \lambda_z^{(u)}+\lambda_z^{(l)}$. 
    }
    \label{fig:SOC_hopping}
\end{figure}

The second effect concerns complex hopping between sites. The GKA rules can be modified to treat complex hoppings. For bonds with inversion symmetry, contributions at order $t\lambda$ in perturbation theory vanish precisely. 
Interactions arising at order $(\lambda)^2$ take the form $J_\mu (2S_l^\mu S_m^\mu -\mathbf{S}_l\cdot\mathbf{S}_m)$, where $\mu$ refers to the direction of  $\vec{\lambda}$. The sign of $J_\mu$ is the same as anticipated from the GKA rules. Thus, spin-dependent hopping between half-filled orbitals results in an antiferromagnetic Ising term plus a ferromagnetic Heisenberg term of half the magnitude. The converse applies to hopping from half-filled orbitals to filled or empty orbitals. 
For all edge-sharing materials, the largest nearest neighbor $\vec{\lambda}_{\alpha \beta}^{lm}$ terms correspond to hopping between the $e_g$ orbitals, which tend to hybridize with the ligands to a much higher degree than the $t_{2g}$ orbitals. As a representative example, there are two relevant sequences for the Z$_1$ bond: $d_{x^2\text{-}y^2}  \ \overrightarrow{_{\text{hop}}}\ p_x \ \overrightarrow{_{L_zS_z}} \ p_y \ \overrightarrow{_{\text{hop}}} \ d_{z^2}$ ($\lambda_z^{(u)}$, via the upper path depicted in Fig.~\ref{fig:SOC_hopping}) and $d_{x^2\text{-}y^2}  \ \overrightarrow{_{\text{hop}}}\ p_y \ \overrightarrow{_{L_zS_z}} \ p_x \ \overrightarrow{_{\text{hop}}} \ d_{z^2}$ ($\lambda_z^{(l)}$, via the lower path). Due to the specific phases of the orbitals, these paths add constructively, $\lambda^z = \lambda_z^{(u)} + \lambda_z^{(l)}$. Within the spin-dependent hopping picture, this manifests into a particularly large $\lambda^z_{(x^2\text{-}y^2,z^2)}$. As discussed previously in Ref.~\onlinecite{stavropoulos2019microscopic}, this may be associated with an antiferromagnetic Kitaev coupling for $d^8$ filling. 

More generally, we expect strong bilinear anisotropic terms in materials with half-filled $e_g$ orbitals, provided
the ligands are sufficiently heavy. This applies most readily to $d^8$ filling, since materials with $d^5$ filling have reduced $p$-$d$ hybridization. In contrast, the $d^3$ materials, with empty $e_g$ orbitals, are expected to exhibit much weaker anisotropic exchange. The above trends and mechanisms highlight
that electron filling and bonding geometry play crucial roles in the relative anisotropy of the magnetic couplings. 

\section{Magnetic exchange couplings for monolayer $MX_2$}
\label{sec:MX2_results}

In this section, we present estimates for the magnetic exchange couplings in monolayer 
$MX_2$ by employing  the \textit{ab-initio}-based projED and 4-state methods, and discuss results
with the expected trends introduced above. The 4-state method is based on total energy mapping through non-collinear, magnetic DFT calculations, including SOC. Each magnetic interaction parameter is related to the energies of four distinct magnetic configurations\cite{xiang2013magnetic,li2021spin}. 
The projED method consists of two steps: i) a finite-cluster Hubbard Hamiltonian is constructed based on Wannier projection of a non-magnetic, relativistic band structure calculation; ii) the Hubbard Hamiltonian is solved via exact diagonalization (ED) and the corresponding effective spin Hamiltonian is extracted through projection onto the low-energy subspace\cite{riedl2019abinitio}.
In the results below, we employ a $d$-only basis for projED; as such, the effects of $p$-$d$ hybridization are downfolded into effective $d$-only single particle terms and renormalized Coulomb terms discussed in Section \ref{sec:metalligand}. While this allows to capture the majority of contributions to the magnetic exchange, as recently pointed out for the d$^7$ case\cite{winter2022magnetic}, there is one important contribution omitted, which represents the regular ferromagnetic Goodenough-Kanamori exchange for $90^\circ$ bonds. This consists of processes where a hole from each metal site meets on a single ligand in different orbitals, and interact via Hund's coupling. Perturbative corrections due to this process are mentioned in each section. Further descriptions of both methods and their relative merits are provided in \cref{sec:methods}. 

In \cref{sec:results_Ualt} we list results obtained with alternative $U_{\rm avg}$ and $J_{\rm avg}$ values for all nine investigated materials with both methods. These values were previously used by some of the authors in other works on the Ni$X_2$ materials\cite{amoroso2020spontaneous,song2022experimental}. The comparison to the results below highlights the independence of the qualitative trends obtained by the two employed methods on these input parameters.

\subsection{$d^8$ $S=1$ materials Ni$X_2$}

\begin{table}
    \centering\def\arraystretch{1.1}
    \begin{ruledtabular}
    \begin{tabular}{C|CCC|CCC}
     \text{Ni}X_2 & \multicolumn{3}{c|}{\text{projED}}  & \multicolumn{3}{c}{\text{4-states}} \\
      & \text{NiCl}_2 & \text{NiBr}_2 & \text{NiI}_2  & \text{NiCl}_2 & \text{NiBr}_2 & \text{NiI}_2 \\
    \hline
    J_1  &           -0.7& -0.8& -1.2& -2.9& -3.9& -6.2 \\
    K_1 &                 0.& +0.1& +1.0& 0.& +0.2& +2.2 \\
    \Gamma_1&             0.& 0.& 0.& 0.& 0.& 0. \\
    \Gamma_1^\prime &     0.& 0.& -0.1& 0.& 0.& -0.1 \\
    J_2 &                 0.& 0.& 0.& -0.1& -0.1& -0.2 \\
    J_3 &                 +0.6& +1.0& +1.8& +0.8& +1.8& +4.2 \\
    A_c &                   0.& +0.1& +0.8& 0.& 0.& +0.4
    \end{tabular}
    \end{ruledtabular}
    \caption{Exchange parameters (in meV) for the Ni$X_2$ monolayer structures using the respective $(U_{\rm avg},J_{\rm avg})$ parameters determined with cRPA for each material (see \cref{tab:cRPA}); extracted with the projED method (left) and the 4-states energy mapping method (right). Note that for the projED method $J_1$ should be corrected with $\delta J_1 \sim -2$ to $-5\,$meV to take additional ferromagnetic contributions into account (see \cref{sec:methods}).
    }
    \label{tab:NiX2}
\end{table}

Experimentally, all bulk Ni$X_2$ materials exhibit magnetic long-range order and were subject of intense investigation a few decades ago due to their metamagnetic and multiferroic behavior\cite{lindgard1975spin,day1976optical,tokunaga2011multiferroicity,kurumaji2013magnetoelectric}. For the monolayer structures, previous theoretical analysis by some of the authors indicated a range of possible ground states: from a topological spin lattice\cite{amoroso2020spontaneous}, composed of vortices and anti-bi-vortices to other competing commensurate and incommensurate phases\cite{amoroso2020spontaneous,ni2021giant,amoroso2021tuning}, including a multiferroic spin spiral state\cite{song2022experimental}. 

In Table~\ref{tab:NiX2}, we report the exchange parameters of model \cref{hamiltonian_eqn} as defined in \cref{eq:interaction_tensor} for Ni$X_2$ monolayers. Values are computed using the projED and 4-state methods, employing $U_{\rm avg}$ and $J_{\rm avg}$ values estimated 
via the cRPA approach (Table~\ref{tab:cRPA}). 

In agreement with the expectations addressed in \cref{sec:trends}, we find $J_1$ to be ferromagnetic (i.e. $J_1<0$) and increasing across the series Cl-Br-I due to increased metal-ligand hybridization. 
In $d^8$ materials, with half-filled $e_g$ orbitals, exchange processes with multiple holes on a single ligand lead to significant ferromagnetic corrections to the value of $J_1$ obtained with projED. As detailed in \cref{sec:methods}, we estimate with perturbation theory $\delta J_1 \sim -2$ to $-5$\,meV. 
It should therefore be noted that the ferromagnetic ligand-mediated exchange $\delta J_1$ is for $d^8$ filling the largest contribution to $J_1$. This can be expected because the $e_g$ orbitals hybridize strongly with the $p$-orbitals. The signs and overall magnitudes obtained for the monolayers are compatible with the bulk values; for example, bulk NiCl$_2$ orders ferromagnetically in plane\cite{katsumata1973effect,lindgard1975spin,billerey1980low}, with ESR and neutron scattering experiments estimating $J_{1,\rm bulk} \approx -3.6$ meV. 

In both theoretical approaches, we find that the nearest neighbor anisotropic exchange is well represented by an antiferromagnetic Kitaev term $K_1$, which may be quite substantial compared to $J_1$ in NiBr$_2$ and NiI$_2$. The magnitude of $K_1$ scales with the SOC strength of the ligand, clearly identifying the origin of the interaction as the hopping process depicted in Fig.~\ref{fig:SOC_hopping}. These findings are compatible with previous work on superexchange processes mediated by $p$-orbital anions with strong SOC\cite{stavropoulos2019microscopic}, 
such as the $S=1$ NiI$_2$ or the $S=3/2$ Cr$^{3+}$ monolayer systems~\cite{xu2018interplay,xu2020}. For example, Ref.~\onlinecite{stavropoulos2019microscopic} anticipated significant AFM Kitaev interactions in $d^8$ compounds. In that case a subset of exchange processes were considered, leading to the prediction $K_1 \gtrsim 2|J_1|$. Here, we find that the previously omitted correction $\delta J_1$ somewhat reduces this ratio, but relatively large AFM Kitaev interactions still seem to be realised in NiI$_2$.

For the longer range couplings, a vanishingly small $J_2$ and significant antiferromagnetic $J_3$, \emph{i.e.} $|J_1|>|J_3|\gg|J_2|$, is found in both methods. 
These results can again be understood by considering the dominant hoppings. For the second and third nearest neighbors we discuss the Z$_2$- and Z$_3$-bonds, defined in \cref{fig:structure}(a). The hoppings between sites on a Z$_2$-bond are dominated by $\tilde{t}=t_{(xy,z^2)}$ and $t_2=t_{(xz,yz)}$. For the Z$_3$-bond we find $t_3=t_{(xy,xy)}$, $t_5=t_{(x^2\text{-}y^2,x^2\text{-}y^2)}$, and $t_6=t_{(z^2,z^2)}$ to be the largest. 
For second neighbors, the $d^8$ filling ensures that only the $t_{2g}$-$e_g$ hopping $\tilde{t}$ is relevant, which leads to a weak ferromagnetic interaction. This contribution is expected to be smaller than the nearest neighbor coupling by a factor $\sim(t_{2nn}/t_{nn})^2$, which explains the relative suppression. In contrast, for third neighbors, there are large $e_g$-$e_g$ hoppings that arise from ligand-assisted hopping paths such as $d_{x^2-y^2} \ \overrightarrow{_{\text{hop}}}\ p_x \ \overrightarrow{_{\text{hop}}}\ p_y \ \overrightarrow{_{\text{hop}}}\ d_{x^2-y^2}$ that are enhanced by metal-ligand hybridization and the spatial extent of the ligand $p$-orbitals. This results in large third neighbor $t_5$ and $t_6$, producing significant third neighbor antiferromagnetic couplings, which tend to increase in magnitude across the series Cl-Br-I. In fact, with both projED and 4-state approaches, we find that $J_3/|J_1|$ increases toward a significant contribution across this series. 

Regarding the single-ion anisotropy $A_c$, with both \mbox{projED} and 4-states methods, we generally find $A_c > 0$, with a magnitude that increases with the SOC strength of the ligand. 
This is compatible with bulk trends. For bulk NiCl$_2$, the single-ion anisotropy was determined to be finite, but small, i.e. below 0.01\,meV\cite{lindgard1975spin,billerey1980low}. For bulk NiBr$_2$, SIA was experimentally\cite{regnault1982inelastic} estimated to be larger $A_c \sim 0.08 \pm 0.05$ meV, which is similar to the monolayer projED estimation $A_c \approx 0.1\,$meV. While it is tempting to attribute this effect to an increase in the effective SOC of the metal via $d$-$p$ hybridization, the situation is not so simple, as outlined in \cref{sec:hopping}. 

The classical ground states of the triangular Heisenberg model with isotropic ferromagnetic $J_1$ and competing FM or AFM $J_2, J_3$ have been thoroughly investigated in the past \cite{rastelli1979non}. Our estimated range of parameters are compatible with two possible magnetic ground states for NiX$_2$ monolayers, namely a ferromagnetic or an incommensurate helix with ordering wavevector $q_{2D}=(q,q)$, the latter displaying lower energy when $4J_3\gtrsim \vert J_1\vert+3\vert J_2\vert$. The increasing trend of $J_3/\vert J_1\vert$ across the Cl-Br-I series indicates an enhanced tendency to stabilise incommensurate helimagnetism for heavier ligands, with NiCl$_2$ monolayer being very close to the FM-spiral phase boundary. These trends can be compared to their bulk counterparts, keeping in mind that interlayer interactions may also affect the magnetic properties of the latter.

Bulk NiCl$_2$ orders ferromagnetically in the plane\cite{katsumata1973effect,lindgard1975spin,billerey1980low} with an ordering wavevector $q_{3D}=(0,0,1.5)$ signalling AFM interlayer interaction between the planes. On the other hand, NiBr$_2$ and NiI$_2$ undergo a series of  transitions: upon decreasing temperature, they both enter first a similar state to bulk NiCl$_2$ of antiferromagnetically coupled ferromagnetic sheets\cite{katsumata1969antiferromagnetic,day1976optical,babu2019magnetic,billerey1977neutron,billerey1980magnetic,kuindersma1981magnetic}, but then upon decreasing $T$ further, they enter another, incommensurate spiral phase\cite{katsumata1969antiferromagnetic,day1980incommensurate,adam1980neutron,regnault1982inelastic,day1984inelastic,babu2019magnetic,billerey1977neutron,billerey1980magnetic,kuindersma1981magnetic}. For NiBr$_2$, the spiral displays a wavevector $(q,q,1.5)$ with small $q\sim 0.03$ and spins lying in the plane of the layer\cite{regnault1982inelastic,babu2019magnetic}. Such a phase can be understood in the context of a model with ferromagnetic $J_1$, antiferromagnetic $J_3$, and $A_c > 0$ with $J_3/|J_1| \gtrsim 0.25$\cite{rastelli1979non}, as well as antiferromagnetic interlayer exchange\cite{regnault1982inelastic},
thus demonstrating a significant $J_3 > 0$ in the bulk materials.
These parameter trends are fully consistent with our monolayer calculations.

For bulk NiI$_2$, the spiral develops concomitantly with a structural transition from rhombohedral to monoclinic cell and displays instead a wavevector $(q,0,\sim 1.46)$, with $q\gtrsim 0.1$ and moments oriented in the plane perpendicular to one of the cubic axes ($\sim 35^\circ$ from the plane)\cite{kuindersma1981magnetic}. Such an orientation is difficult to understand without invoking bond-dependent couplings referenced to the cubic axes. Since the ordering vector is perpendicular to a bond, the ordering pattern consists of stripes in which moments linked by one of the nearest neighbor bond types are always aligned in parallel. In the presence of a sizeable antiferromagnetic $K_1>0$, there is an energetic preference for the moments to align in the plane perpendicular to the associated cubic axis of the bond perpendicular to the $q$-vector. This is precisely what is observed experimentally. Thus the particular moment orientation should be taken as evidence for significant $K_1 >0$ in bulk NiI$_2$. 
Albeit the larger ordering wavevector of bulk NiI$_2$ is still consistent with our prediction of increasing magnetic frustration across the Cl-Br-I series, the in-plane component of bulk $q_{3D}$ would suggest, within an isotropic Heisenberg model, a strong antiferromagnetic $J_2$\cite{rastelli1979non,kuindersma1981magnetic}. Such a significant deviation from expected and estimated trends for magnetic interactions in the NiX$_2$ class is difficult to rationalize, and one may wonder how appropriate it is to compare bulk and monolayer NiI$_2$.
A recent experimental analysis of NiI$_2$ magnetic properties down to the monolayer limit via complementary optical techniques could not resolve the 2D ordering wavevector, both $q_{2D}=(q,q)$ and $(q,0)$ being compatible with the detected symmetries\cite{song2022experimental}. At the same time, a quite strong AFM interlayer interaction, $\sim 0.45 \vert J_1\vert$, was deduced from the layer-dependent spiral transition temperature\cite{song2022experimental}, much larger than the interlayer exchange of NiBr$_2$, $\sim 0.1\vert J_1\vert$\cite{regnault1982inelastic}. These facts suggest that other mechanisms, not included in the 2D model considered here, may play an important role in shaping the bulk magnetic properties of NiI$_2$.

Finally, we note that a large nearest neighbor biquadratic exchange $B\,(\mathbf{S}_i\cdot \mathbf{S}_j)^2$ with $B<0$ was recently invoked to explain the collinear ground state of NiCl$_2$, compared to non-collinear helimagnetic ground states of NiBr$2$ and NiI$_2$. To test the possibility of large $B$, we also computed all higher order couplings using projED for the monolayer structure, which is capable of capturing 4-spin interactions\cite{riedl2021spin}. However, we find no 4-spin terms exceeding 0.01\,meV. This result can be understood from the fact that $B$ arises at order $t^4/U^3$ in perturbation theory, and is generally expected to be significant for nearly itinerant electrons. Nonetheless, it could still be relevant for systems like NiCl$_2$ that are at the verge of the FM-spiral phase transition and, as such, sensitive to other weak interactions not included in the model Eq. (\ref{hamiltonian_eqn}).

\subsection{$d^3$ $S=3/2$ materials V$X_2$}

For V$X_2$ compounds, the exchange parameters extracted with both \textit{ab-initio} methods are given in \cref{tab:VX2}. 
Good agreement is found both in magnitude and trends between 4-state and projED approaches. 

We find antiferromagnetic (AFM) nearest neighbor Heisenberg coupling $J_1$, with a decreasing magnitude as a function of the ligand atomic number. As outlined in \cref{sec:trends}, the dominant hopping processes suggest a competition between antiferromagnetic and ferromagnetic contributions in V$X_2$, resulting in a net antiferromagnetic interaction; 
the ligand-assisted hoppings become more important for the heavier ligands, leading to a stronger contribution of the ferromagnetic exchange and resulting in the overall decrease of $|J_1|$ across the series Cl-Br-I. 

\begin{table}
    \centering\def\arraystretch{1.1}
    \begin{ruledtabular}
    \begin{tabular}{C|CCC|CCC}
     \text{V}X_2 & \multicolumn{3}{c|}{\text{projED}}  & \multicolumn{3}{c}{\text{4-states}} \\
      & \text{VCl}_2 & \text{VBr}_2 & \text{VI}_2  & \text{VCl}_2 & \text{VBr}_2 & \text{VI}_2 \\
    \hline
    J_1 &           +4.9& +3.6& +2.6& +4.2& +2.5& +0.6 \\
    K_1 &                 0.& 0.& 0.& 0.& 0.& -0.1 \\
    \Gamma_1&             0.& 0.& 0.& 0.& 0.& +0.1 \\
    \Gamma_1^\prime&     0.& 0.& 0.& 0.& 0.& 0. \\
    J_2 &                 0.& +0.1& +0.1& 0.& 0.& +0.1 \\
    J_3 &                 0.& +0.1& +0.1& 0.& +0.1& +0.2 \\
    A_c &                   0.& 0.& 0.& 0.& 0.& 0.
    \end{tabular}
    \end{ruledtabular}
    \caption{Exchange parameters (in meV) for the V$X_2$ monolayer structures using the respective $(U_{\rm avg},J_{\rm avg})$ parameters determined with cRPA for each material (see \cref{tab:cRPA}); extracted with the projED method (left) and the 4-states energy mapping method (right). 
    }
    \label{tab:VX2}
\end{table}

The second and third nearest neighbor couplings, $J_2$ and $J_3$ respectively, turn out to be small generally. This can be understood considering the primary third neighbor hopping occurs between $e_g$ orbitals via paths like $e_g  \ \overrightarrow{_{\text{hop}}}\ p  \ \overrightarrow{_{\text{hop}}}\ p  \ \overrightarrow{_{\text{hop}}}\ e_g$. For $d^3$ filling, such hoppings are irrelevant to the magnetic couplings. As a result, the long-range interactions are suppressed compared to $d^8$ filling. Similarly, consistent with the discussion in \cref{sec:trends}, the bilinear anisotropic exchange is negligible, since the ligand SOC only makes relevant corrections to hopping between $e_g$ orbitals. 

Our results show consistency with behaviors experimentally observed in the VX$_2$ bulk systems. 
Estimated intralayer nearest-neighbor exchange couplings fitting data from susceptibility \cite{niel1977magnetic} and INS data\cite{kadowaki1987experimental,kadowaki1985neutron}, were reported
to be $J_1 = +3.8$\,meV and $J_1 = +2.8$\,meV, for bulk VCl$_2$ and VBr$_2$, which supports the {\it ab-initio} trends.  Moreover, the AFM N\'eel 120$^\circ$ spin order was observed, 
with transition temperatures of $T_{\rm N}= 36\,$K and 29\,K, for the two systems respectively\cite{hirakawa1983study,nishi1984neutron,kadowaki1985neutron,kadowaki1987experimental}. 
In accordance with our {\it ab-initio} intralayer estimates, these results suggest that monolayer VCl$_2$ and VBr$_2$ can be sufficiently described by short-range AFM interactions. Given the absence of anisotropic couplings, we would expect these ordering temperatures to be significantly reduced in monolayers. However the ordering pattern is not expected to differ. 

At variance, the physics of VI$_2$ appears less obvious. Bulk VI$_2$ displays AFM 120$^\circ$ correlations at high temperature, but ultimately orders at $T_{\rm N_2} = 14.4\,$K into an AFM collinear zigzag stripe order\cite{kuindersma1979magnetic,hirakawa1983study}. 
No experimental estimates of $J_1$ are available (to the best of our knowledge). 
However, the lower $T_{\rm N}$ values with respect to VCl$_2$ and VBr$_2$ suggest reliability of our {\it ab-initio} estimates: 
the former suggests energetic competition between AFM phases that could be ascribed to the non-zero intralayer $J_3$ contribution estimated in monolayer VI$_2$; 
the latter supports the argued $J_1$ decrease across the series Cl-Br-I (Table~\ref{tab:VX2}).

\subsection{$S=5/2$ materials Mn$X_2$}

As discussed in \cref{sec:trends}, the magnetic exchange interactions in Mn$X_2$ are expected to be relatively suppressed as a consequence of the large energy cost for transfer of electrons between metal sites. This is indeed what is found with both projED and 4-state approaches, as detailed in Table~\ref{tab:MnX2}. 
Due to the partially filled $e_g$ orbitals, $J_1$ obtained with projED has to be corrected, with $\delta J_1 \sim -0.4$ to $-0.9$\,meV (see \cref{sec:methods}).
The nearest-neighbor Heisenberg $J_1$ exchange is found to be antiferromagnetic, with little variation with ligand elements, in contrast to the behaviors discussed for Ni$X_2$ and V$X_2$. Anisotropic exchange associated with the $e_g  \ \overrightarrow{_{\text{hop}}}\ p  \ \overrightarrow{_{\text{hop}}}\ p  \ \overrightarrow{_{\text{hop}}}\ e_g$ hopping processes should be finite, but it represents only a small fraction of the total contributions, and is therefore negligible. With both approaches we find longer range couplings $J_2$ and $J_3$ in the range 0.01 to 0.1\,meV, which may be significant given the relative suppression of $J_1$.

\begin{table}
    \centering\def\arraystretch{1.1}
    \begin{ruledtabular}
    \begin{tabular}{C|CCC|CCC}
     \text{Mn}X_2 & \multicolumn{3}{c|}{\text{projED}}  & \multicolumn{3}{c}{\text{4-states}} \\
      & \text{MnCl}_2 & \text{MnBr}_2 & \text{MnI}_2  & \text{MnCl}_2 & \text{MnBr}_2 & \text{MnI}_2 \\
    \hline 
    J_1 &           +0.8& +0.7& +0.7& +0.1& +0.1& +0.1 \\
    K_1 &                 0.& 0.& 0.& 0.& 0.& 0. \\
    \Gamma_1&             0.& 0.& 0.& 0.& 0.& 0. \\
    \Gamma_1^\prime &     0.& 0.& 0.& 0.& 0.& 0. \\
    J_2 &                 0.& 0.& 0.& 0.& 0.& 0. \\ 
    J_3 &                 0.& +0.1& +0.1& 0.& 0.& +0.1 \\
    A_c &                   0.& 0.& 0.& 0.& 0.& 0.
    \end{tabular}
    \end{ruledtabular}
    \caption{Exchange parameters (in meV) for the Mn$X_2$ monolayer structures using the respective $(U_{\rm avg},J_{\rm avg})$ parameters determined with cRPA for each material (see \cref{tab:cRPA}); extracted with the projED method (left) and the 4-states energy mapping method (right). 
    Note that for the projED method $J_1$ should be corrected with $\delta J_1 \sim -0.4$ to $-0.9$\,meV to take additional ferromagnetic contributions into account (see \cref{sec:methods}).
    }
    \label{tab:MnX2}
\end{table}

Consistently with the nearly suppressed magnetic exchange interactions, very low experimental transition temperatures are reported for the bulk Mn$X_2$ systems. 
Particularly, bulk MnCl$_2$ and MnBr$_2$ exhibit two successive transitions at $T_{\rm N_1} = 1.96\,$K, $T_{\rm N_2} = 1.81\,$K\cite{osti_4344362} and $T_{\rm N_1} = 2.32\,$K, $T_{\rm N_2} = 2.17\,$K\cite{sato1995successive,iio1990successive,sato1993neutron} respectively. In both cases,  the intermediate phase has been identified as an incommensurate phase, and the low-temperature structure as a double stripe order phase. 
Bulk MnI$_2$ exhibits even three subsequent magnetic transitions below 4\,K\cite{sato1993neutron,sato1995successive}. The lowest temperature phase was identified as a spiral order with an incommensurate wavevector\cite{cable1962neutron}. Given the very small magnitude of the exchange couplings and large $S = 5/2$, we agree with previous speculations\cite{sato1993neutron,utesov2017cascades} that long-range dipolar interactions may ultimately be relevant for both, bulk and monolayer systems.

\section{Conclusions}
\label{sec:conclusion}

With this study we investigated the effective magnetic models
in the monolayer family of triangular lattice materials $MX_2$ ($M$=\{V, Mn, Ni\}, $X$=\{Cl, Br, I\}), where spin-orbit coupling effects are dominantly caused by the ligand element. We also took the opportunity of such a larger study to benchmark two \textit{ab-initio} methods used to extract effective spin models for real materials.

Considering the very different nature of the two \textit{ab-initio}-based  methods considered here, the results agree remarkably well in terms of signs and trends in the relative magnitudes. The points of disagreement are typically the overall magnitude. Among others, this may be assigned to the implementation of the parameter values $U_{\rm avg}$ and $J_{\rm avg}$, since they enter either in the Hubbard Hamiltonian for projED, or in the DFT+U framework for the 4-states method. For consistency, we compared both methods using the same parameter values in spite of the fact that this may not be the ideal parameter set especially for the 4-states method. On the other hand, in the 4-states method an overall linear dependence  varying these parameters was found (compare \cref{tab:MX2_Ualt}) while the projED method relies stronger on choosing appropriate parameter sets such as the ones used in this work, obtained with cRPA. In the case of half-filled $e_g$ orbitals for the projED method ferromagnetic contributions from additional processes have to be considered, which we estimated here via perturbation theory.

Regarding the magnetic parameters for the triangular lattice compounds in general, we find trends for the monolayer structures consistent with expectations based on experimental observations in the bulk case. Noticeably, for monolayer NiI$_2$  we do find  sizeable Kitaev coupling suggested in previous works\cite{stavropoulos2019microscopic,amoroso2020spontaneous}, as well as large ferromagnetic $J_1$ and significant antiferromagnetic $J_3$ isotropic couplings. The V$X_2$ monolayer structures can all be described well in the framework of nearest neighbor AFM Heisenberg models, and the Mn$X_2$ materials reveal overall suppressed magnetic exchange, hinting at long-range dipolar interactions as potentially the most important factor for the magnetic ground state.

Finally, already the use of simplified models made it possible to understand the underlying mechanism of the magnetic exchange in the triangular $MX_2$ compounds. Consideration of two dominant hopping parameters enabled us to identify the \mbox{(anti-)}ferromagnetic nature of $J_1$ in the ($d^3$) $d^8$ materials, including the trends of additional or competing ferromagnetic contributions for larger ligands. 
Our analysis further allows for a physical insight in the peculiar situation of $d^8$ materials, displaying i) significant enhancement of the third-nearest-neighbor exchange interaction $J_3$, especially when compared to shorter-range $J_2$, and ii) substantial anisotropic exchange/Kitaev interactions, increasing with ligand atomic number. The first effect can be understood only if the specific hopping interactions and involved orbital states contributing to the exchange interactions are properly taken into account: as the hopping processes relevant for $J_2$ and $J_3$ are fundamentally different, the conventional expectation that exchange interaction should roughly scale with the inverse of the bond length is, at best, inaccurate. Additionally, $J_3$ is dominated by ligand-assisted hopping processes, and as such can be directly tuned by the metal-ligand hybridization (hence by appropriate choices of the ligand). Similarly, relativistic anisotropic exchanges arise mostly from ligands spin-orbit coupling, which tunes through ligand-assisted hoppings the effective SOC between the magnetic ions and strongly modifies its matrix elements as compared to the atomic limit. It follows that the resulting strength of the anisotropic exchange found in $d^8$ materials is mostly driven by the atomic SOC of the ligands, as well as by the metal-ligand hybridization again. The emerging picture thus possibly suggests strategies to enforce magnetic frustation and Kitaev-type interactions, both sought for in the quest of exotic magnetic phases, alternative to existing ones based mostly on suitable choices of heavier ($4d$ or $5d$) transition metals.

\begin{acknowledgments}
R.V. and K.R. acknowledge support by the Deutsche Forschungsgemeinschaft (DFG, German Research Foundation) for funding through TRR 288 --- 422213477 (projects A05 and B05).
 D.A. and S.P. acknowledge support by the Nanoscience Foundries and Fine Analysis (NFFA-MIUR Italy) project. P.B. and S.P. acknowledge financial support from the Italian Ministry for Research and Education through PRIN-2017 projects ‘Tuning and understanding Quantum phases in 2D materials—Quantum 2D’ (IT-MIUR grant No. 2017Z8TS5B) and ‘TWEET: Towards ferroelectricity in two dimensions’ (IT-MIUR grant No. 2017YCTB59), respectively. D.A., P.B. and S.P. also acknowledge high-performance computing systems operated by CINECA (IsC722DFmF, IsC80-Em2DvdWd, IsC88-FeCoSMO and IsB21-IRVISH projects). D.A. is grateful to M. Verstraete and B. Dup\'e (ULiege) for the time allowed to work on the writing of this paper.
\end{acknowledgments}

\appendix

\section{Relation between exchange values in crystallographic and cubic coordinate systems}
\label{sec:coordinates}

The parameterization convention used in the main text follows the extended Heisenberg-Kitaev model with focus on the bond-dependent Ising-like Kitaev interaction. With this convention it is easy to identify such general bond-dependent parameters. For example, on a Z$_1$-bond between sites $l$ and $m$, the magnetic exchange matrix is defined as follows:
\begin{align} \label{eq:para_cubic}
\mathbb{J}_{lm}^{\text{Z}_1}= \begin{pmatrix} J & \Gamma & \Gamma^\prime \\ \Gamma & J & \Gamma^\prime \\ \Gamma^\prime & \Gamma^\prime & J+K \end{pmatrix}.
\end{align}
while on an X$_1$-bond it can be determined by $C_3$ rotation about the out-of-plane axis arriving at:
\begin{align}
\mathbb{J}_{lm}^{\text{X}_1}= \begin{pmatrix} J+K & \Gamma^\prime & \Gamma^\prime \\ \Gamma^\prime & J & \Gamma \\ \Gamma^\prime & \Gamma & J \end{pmatrix}.
\end{align}

Alternatively, the interactions can also be expressed in a coordinate system oriented along the crystal axes $(ab^\ast c)$, where $b^\ast$ is perpendicular to $a$ and $c$. 
By symmetry, the magnetic exchange on a bond along the $a$ axis (e.g. Z$_1$-bond) can be parametrized with:
\begin{align}
    \mathbb{J}_{lm}^{\text{Z}_1} = \begin{pmatrix}
    J_{aa} & 0 & 0 \\
    0 & J_{b^\ast b^\ast} & J_{b^\ast c} \\
    0 & J_{b^\ast c} & J_{cc}
    \end{pmatrix}.
    \label{eq:tensor_dan}
\end{align}
The corresponding exchange matrix on the X$_1$-bond can then be again obtained by $C_3$ rotation about the $c$ axis:
\begin{align}
    \mathbb{J}_{lm}^{\text{X}_1} = \begin{pmatrix}
    \frac{1}{4}(J_{aa}+3J_{b^\ast b^\ast}) & -\frac{\sqrt{3}}{4}(J_{aa}-J_{b^\ast b^\ast}) & -\frac{\sqrt{3}}{2}J_{b^\ast c} \\
    -\frac{\sqrt{3}}{4}(J_{aa}-J_{b^\ast b^\ast}) & \frac{1}{4}(3J_{aa}+J_{b^\ast b^\ast} & -\frac{1}{2}J_{b^\ast c} \\
    -\frac{\sqrt{3}}{2}J_{b^\ast c} & -\frac{1}{2}J_{b^\ast c} & J_{cc}
    \end{pmatrix}.
    \label{eq:tensor_120_dan}
\end{align}

Both approaches are equivalent and can be directly translated into each other via the following relations:

\begin{align}
    J &= \frac{1}{6} \left(3 J_{aa} + J_{b^\ast b^\ast} + 2 J_{cc} + 2\sqrt{2} J_{b^\ast c} \right) \\
    K &= \frac{1}{2} \left( - J_{aa} + J_{b^\ast b^\ast} - 2 \sqrt{2} J_{b^\ast c} \right) \\
    \Gamma &= \frac{1}{6} \left(-3 J_{aa} + J_{b^\ast b^\ast} + 2 J_{cc} + 2\sqrt{2} J_{b^\ast c} \right) \\
    \Gamma^\prime &= \frac{1}{6} \left(-2 J_{b^\ast b^\ast} + 2 J_{cc} - \sqrt{2} J_{b^\ast c} \right)
\end{align}

Independent of the coordinate system, the exchange matrix can conventionally be decomposed into three distinct contributions. The fully isotropic part with respect to spin orientations ---in contrast to bond anisotropy--- corresponds to $J_{lm}^{\text{iso}}=\frac{1}{3}\text{Tr}(\mathbb{J}_{lm})$. In the case of the cubic coordinate system with \cref{eq:para_cubic}, this would correspond to $J_{lm}^{\text{iso}}=J+K/3$. The anisotropic terms with respect to spin orientation are composed of the symmetric term,  $\mathbb{J}_{lm}^{\text{S}}=\frac{1}{2}(\mathbb{J}_{lm}+\mathbb{J}_{ml})-J_{lm}^{\text{iso}}\,\rm{I}$ and the anti-symmetric term $\mathbb{J}_{lm}^{\text{A}}=\frac{1}{2}(\mathbb{J}_{lm}-\mathbb{J}_{ml})$. The latter vanishes in the presence of an inversion center in the middle of the bond, as it is the case for the materials investigated in this work.

\section{Methods} 
\label{sec:methods}

\subsection{Monolayer $MX_2$ structures}

Within the projED and 4-state methods, calculations have been performed on fixed triangular $MX_2$ monolayer structures, 
which have been fully relaxed through structural relaxations as implemented in VASP\cite{kresse_paw}, 
within the generalized gradient approximation (GGA) and Perdew–Burke–Erzenhof (PBE)\cite{pbe} functional. 
Particularly, no Hubbard-$U$ and SOC corrections were employed for all the structures' optimization, whereas different magnetic orderings were considered according to calculated energetics 
(i.e. lowest energy configuration without SOC contribution): FM order for Ni$X_2$ systems, consistent with previous works\cite{amoroso2020spontaneous} and
120$^\circ$ AFM order for Mn$X_2$ and V$X_2$ systems. The obtained in-plane lattice parameters ($|{\bm a}|=|{\bm b} |$) are reported in Table~\ref{tab:latt_param}; 
the {\bf c} length was fixed to 20.8~\AA, to insert a vacuum distance between periodic repetition of the layers along this direction.  

\begin{table}[h!]
    \centering
    \begin{ruledtabular} 
    \begin{tabular}{ccc|ccc|ccc} 
    & calc. & exp.\cite{mcguire_rev} & & calc. & exp.\cite{mcguire_rev} & & calc. & exp.\cite{mcguire_rev} \\
NiCl$_2$ & 3.49 & 3.48  & VCl$_2$ &  3.54 & 3.6 &  MnCl$_2$ & 3.70 & 3.71 \\
NiBr$_2$ & 3.69 & 3.7 & VBr$_2$ & 3.75  & 3.77 &  MnBr$_2$ & 3.88 & 3.89 \\
NiI$_2$ & 3.96  & 3.9  & VI$_2$  & 4.05 & 4.06  &  MnI$_2$  & 4.15 & 4.15 \\
\end{tabular}
    \end{ruledtabular}
    \caption{In-plane lattice parameters in angstrom (\AA) for the studied triangular $MX_2$ monolayer systems. Values obtained from DFT-calculations are compared with corresponding experimental data for bulk structures. }
    \label{tab:latt_param}
\end{table}

\subsection{Constrained RPA}

We estimated the electronic two-particle interaction terms with the constrained random-phase approximation (cRPA)~\cite{Aryasetiawan2004,Aryasetiawan2006}, as implemented in the FHI-gap code~\cite{fhigap}. The approach is similar to a recent cRPA study\cite{yekta2021strength}, applied here for the discussed relaxed monolayer structures and considering the full five $3d$ orbitals of the metal elements in each case. 
The atomic-like spherical symmetric expressions for $d$ block electrons derived from Slater integrals~\cite{liechtenstein1995density} $F_k$ can be obtained via the following relations (for orbitals $\alpha$, $\beta$ and angular momentum quantum number $l$):
\begin{align}
U_{\rm avg}&=\frac{1}{(2l+1)^2} \sum_{\alpha \beta} U_{\alpha \beta}=F_0, \\
J_{\rm avg} &= \frac{7}{5}\, \frac{1}{2l(2l+1)} \sum_{\alpha \neq \beta} J_{\alpha \beta}=\frac{F_2+F_4}{14}.
\end{align}

\subsection{ProjED}

The projED method~\cite{riedl2019abinitio} consists of two main steps.
First, an effective electronic multi-orbital Hubbard Hamiltonian $\mathcal{H}_{\text{tot}}$ is determined:
\begin{align}
\mathcal{H}_{\rm tot} &= \mathcal{H}_{\rm hop}+ \mathcal{H}_{\rm U}, \label{eq:Htot}
\\
\mathcal{H}_{\rm hop} &= \sum_{lm\,\alpha \beta} \sum_{\sigma \sigma^\prime} t_{l\alpha,m \beta}^{\sigma \sigma^\prime}\, c_{l\alpha\sigma}^\dagger c_{m\beta \sigma^\prime}, \\
\mathcal{H}_{\rm U} &= \sum_{l\, \alpha \beta \gamma \delta} \sum_{\sigma \sigma^\prime} U_{l \alpha \beta \gamma \delta}^{\sigma \sigma^\prime}\, c_{l \alpha \sigma}^\dagger c_{l \beta \sigma^\prime}^\dagger c_{l \delta \sigma^\prime} c_{l \gamma \sigma},
\end{align}
where $\{l,m\}$ are site, $\{\alpha,\beta\}$ orbital and $\{\sigma,\sigma^\prime\}$ spin indices. $\mathcal{H}_{\rm hop}$ describes non-magnetic hopping processes between the $3d$ electronic (spin-)orbitals of the transition metals $M=$\{Ni, V, Mn\} and $\mathcal{H}_{\rm U}$ the respective two-particle Coulomb interaction. 
The hopping parameters were computed \textit{ab-initio} using the Full Potential Local Orbital~\cite{fplo} (FPLO) code within the generalized gradient approximation~\cite{pbe} (GGA) for each monolayer structure. The hopping parameters are extracted via Wannier projection based on full relativistic non-spin-polarized band structures. This allows to effectively consider the influence of the heavy ligands involved, especially in the case of the $M$I$_2$ series, and results in \textit{complex} hopping parameters.
To ensure consistent treatment of the two-particle Coulomb interaction within the investigated series, the Coulomb parameters were extracted for each material within the cRPA method, as described above. 

In a second step, an effective spin Hamiltonian is extracted:
\begin{align}
    \mathcal{H}_{\rm eff} = \mathbb{P} \mathcal{H}_{\rm tot}\mathbb{P} = \sum_{l\mu m \nu} J_{lm}^{\mu \nu} S_l^\mu S_m^\nu
\end{align}
with $\mu$, $\nu \in \{x,y,z\}$. The projection operator $\mathbb{P}$ projects the electronic Hamiltonian onto its low-energy subspace and then maps onto the respective spin operator representation. The low-energy eigenstates, required for the first step, are computed via exact diagonalization of the five-orbital electronic Hamiltonian on finite two-site clusters. After reducing the Hilbert space onto an energetic regime where spin degrees of freedom are sufficient to describe the relevant physical properties, the Hamiltonian can be mapped from the electronic onto the spin picture using Stevens operators. In case of $S=1/2$ this would correspond to the well-known Pauli matrices.

Estimating exchange parameters in the above introduced $d$-only basis, ferromagnetic corrections arising from exchange processes with multiple holes on a single ligand are omitted. These contributions are dominantly relevant for the isotropic interaction in materials with partially filled $e_g$ orbitals, since $p$ orbitals hybridize predominantly with $e_g$ orbitals.
We estimate $\delta J_1$ for the relevant fillings via perturbation theory.

For $d^8$ filling, the ferromagnetic correction to the \mbox{projED} results can then be estimated by:
\begin{align} \label{eq:deltaJ_Ni}
    \delta J_1 = -\frac{1}{S^2} \frac{(t_{pd}^\sigma)^4 J_{\rm H}^p}{\Delta_{pd}^2(\Delta_{pd}+U_p/2)^2}
\end{align}
where $\Delta_{pd}$ is the charge transfer energy from $d$ to $p$ orbitals, $U_p$ is the excess ligand Coulomb repulsion, $J_{\rm H}^p$ is the ligand Hund's coupling, and $t_{pd}^\sigma$ is the $e_g$-$p$ hopping integral in the Slater-Koster scheme. On the basis of non-relativistic band-structure calculations, we estimate $t_{pd}^\sigma \sim 1$\,eV and $\Delta_{pd} \sim 3$ to $4\,$eV for Ni$X_2$, while we take $U_p \sim 4$\,eV, $J_H^p \sim 0.3\,U_p$ following Ref.~\onlinecite{liu2018pseudospin}.

For $d^3$ filling, the ferromagnetic correction can be approximated by:
\begin{align} \label{eq:deltaJ_V}
    \delta J_1 = -\frac{3}{S^2} \frac{(t_{pd}^\pi)^4 J_{\rm H}^p}{\Delta_{pd}^2(\Delta_{pd}+U_p/2)^2}
\end{align}
where we estimate the $t_{2g}$-$p$ hopping $t_{pd}^\pi \sim$ 0.7\,eV, and $\Delta_{pd} \sim 4$ to 5\,eV. Again, we may take $U_p \sim 4$\,eV, $J_{\rm H}^p \sim 0.3\,U_p$ following Ref.~\onlinecite{liu2018pseudospin}. This provides an estimate of $\delta J_1 \sim -0.3$ to $-0.7$\,meV. The correction is significantly reduced compared to the Ni$X_2$ compounds due to the weaker $t_{2g}$-$p$ hybridization. As a result, the couplings are well approximated by the $d$-only terms.

For $d^5$ filling, $\delta J_1$ may be estimated via: 
\begin{align} \label{eq:deltaJ_Mn}
    \delta J_1 = -\frac{1}{S^2} \frac{[(t_{pd}^\sigma)^4+2(t_{pd}^\sigma)^2(t_{pd}^\pi)^2+3(t_{pd}^\pi)^4] J_{\rm H}^p}{\Delta_{pd}^2(\Delta_{pd}+U_p/2)^2}.
\end{align}
Taking $t_{pd}^\sigma \sim 1$\,eV, $t_{pd}^\pi \sim 0.7$\,eV, $\Delta_{pd} \sim 4$ to $5$\,eV, $U_p \sim 4$\,eV, and $J_{\rm H}^p \sim 0.3\,U_p$\cite{liu2018pseudospin} yields an estimate of the correction to $\delta J_1 \sim -0.4$ to $-0.9$\,meV.

Extracting the local single-ion anisotropy, we considered a linked-cluster expansion including two-site clusters in order to take contributions from nearest neighbor interactions next to the local contributions into account.

For further neighbor interactions, two-site clusters are used with hopping parameters between the relevant sites. Note that this procedure neglects contributions through indirect hopping paths that involve more than two magnetic sites, but is unavoidable for fillings approaching half-filling due to the computationally demanding exact diagonalization step. Especially for long-range interactions such as $J_3$ this approximation may 
be cautiously examined. For the Ni$X_2$ series we tested the neglected contributions via linked cluster expansion, as used previously with projED\cite{winter2017importance}, including up to three magnetic sites. At least for this class of materials we find only minor modifications below 1\%.

\subsection{4-state total energies}

The estimate of the magnetic interaction parameters employing the ``4-state method'' relies on DFT calculations of the total energies associated to 
various spin configurations to be mapped on the classical spin Hamiltonian written in \cref{hamiltonian_eqn}. This approach is known as energy-mapping analysis. 
Specifically, the 4-state energy mapping method, which is explained in detail in Refs \onlinecite{xiang2013magnetic,xu2018interplay}, 
allows us to extract the magnetic exchange interaction, both the isotropic and anisotropic contributions, between a selected pair of magnetic sites by performing 
DFT plus SOC energy calculations on four ordered noncollinear spin states. 
It is based on the use of supercells, which allow to exclude couplings with unwanted distant neighbors. The method is however tied to the choice of  
DFT basis, implementation of exchange correlation functional used in DFT and to specific computational parameters (i.e. k-points sampling, Hubbard U corrections within DFT+U, etc.). 

In this work, we performed calculations of the SIA, first- and second-neighbors interaction using a $5\times4\times1$ supercell, while a $6\times3\times1$ supercell was used for the estimate of the third-neighbor interaction. Such large cells should exclude a significant influence from next neighbors. 
We built supercells from the periodic repetition of the $MX_2$ monolayer unit cell; structural details are reported in Sec.~\ref{sec:MX2_results}.

By means of this method we can
obtain all the elements of the exchange tensor for a chosen magnetic pair, thus gaining direct access to the symmetric anisotropic exchange part (or
two-ion anisotropy) and the antisymmetric anisotropic part (the DM interaction) of the full exchange. 
In particular, we performed direct calculations on the magnetic $M$-$M$ pairs parallel to the crystallographic $a$ direction (Fig.~\ref{fig:structure}), 
determining the exchange tensor reported in \cref{eq:tensor_dan}. 
The interaction between the five other nearest-neighbor pairs can be evaluated via the three-fold rotational symmetry, as in \cref{eq:tensor_120_dan}.
In all our systems, the tensor turned out to be symmetric or, equivalently, any anti-symmetric (DM-like) contribution was found to be negligible.

\section{Details of non-spin-polarized hopping parameters}
\label{sec:hopping}

To understand the microscopic mechanisms dictated by symmetry considerations of the space group $P\bar{3}m1$ (164) it is insightful to first consider the underlying electronic processes, which we model with an effective tight-binding Hamiltonian reduced on the magnetic ions $M$. The corresponding Hamiltonian is given in \cref{eq:Htot}. The hopping parameters discussed in this section were obtained using the Full Potential Local Orbital~\cite{fplo} (FPLO) code within the generalized gradient approximation~\cite{pbe} (GGA) and served in the relativistic case as basis for the projED results in the main text.

\subsubsection{Effects without spin-orbit coupling}

We first consider the case without spin-orbit coupling effects. In this case $t_{l\alpha,m\beta}^{\sigma \sigma^\prime}=t_{l\alpha,m\beta}^{\sigma \sigma^\prime} \delta_{\sigma \sigma^\prime}$. 

\begin{table}
    \centering\def\arraystretch{1.1}
    \begin{ruledtabular}
    \begin{tabular}{R|RRRRR|RRRRR}
    \multicolumn{1}{C}{} & \multicolumn{5}{C}{\text{generic Z-bond}} & \multicolumn{5}{C}{\text{Z}_1\text{-bond VCl}_2} \\
    & d_{xy} & d_{xz} & d_{yz} & d_{x^2\text{-}y^2} & d_{z^2}  & d_{xy} & d_{xz} & d_{yz} & d_{x^2\text{-}y^2} & d_{z^2}\\
    \hline
    d_{xy}  & t_3&		 t_4&		 t_4&		 0&		\tilde{t}&
    \mathbf{-248}&		 10&		 10&		 0&		 \mathbf{121}\\
    d_{xz}  & t_4&		t_1&		 t_2&		 t_8&		 t_7&  
    10&		 68&		 30&		 0&		-3 \\
    d_{yz}  &    t_4&		 t_2&		t_1&		-t_8&		 t_7&
    10&		 30&		 68&		0&		-3 \\
    d_{x^2\text{-}y^2} &  0&		 -t_8&		t_8&		 t_5&		 0&
    0&		 0&		0&		-90&		 0 \\
     d_{z^2} & \tilde{t}&		 t_7&		 t_7&		 0&		 t_6 &
     \mathbf{121}&  -3&		-3&		 0&		-2 \\
    \hline
    \hline
    \multicolumn{1}{C}{} & \multicolumn{5}{C}{\text{Z}_2\text{-bond VCl}_2} & \multicolumn{5}{C}{\text{Z}_3\text{-bond VCl}_2} \\
    & d_{xy} & d_{xz} & d_{yz} & d_{x^2\text{-}y^2} & d_{z^2}  & d_{xy} & d_{xz} & d_{yz} & d_{x^2\text{-}y^2} & d_{z^2}\\
    \hline
    d_{xy}  & 1&		-2&		-2&		 0&		 \mathbf{16}&
            \mathbf{-19}&		 2&		 2&		 0&		 13\\
    d_{xz}  &   -2&		 5&		\mathbf{-13}&		 2&		 2&    
            2&		 2&		-4&		-2&		-2\\
    d_{yz}  &   -2&		\mathbf{-13}&		 5&		-2&		 2&    
            2&		-4&		 2&		 2&		-2\\
    d_{x^2\text{-}y^2} & 0&		 2&		-2&		-1&		 0&
             0&		-2&		 2&		 \mathbf{56}&		 0 \\
     d_{z^2} & \mathbf{16}&		 2&		 2&		 0&		 1&         
             13&		-2&		-2&		 0&		\mathbf{-20}
    \end{tabular}
    \end{ruledtabular}
    \caption{Non-relativistic, non-spin-polarized hopping parameters for a generic Z-bond, considering only symmetry restrictions, as well as hopping parameters (in meV) for VCl$_2$ on a Z$_1$-, Z$_2$-, and Z$_3$-bond as defined in \cref{fig:structure}. The dominant hoppings are highlighted for Z$_1$ ($t_3$, $\tilde{t}$), Z$_2$ ($t_2$, $\tilde{t}$) and Z$_3$ ($t_3$, $t_5$, $t_6$).
    }
    \label{tab:hoppings}
\end{table}

Let us consider the symmetry restrictions on first, second and third neighbor bonds, which can be all described by the same reduced matrices due to $2/m$ symmetry at their bond center. 
For this discussion, we focus on Z$_1$-, Z$_2$ and Z$_3$-bonds, which can be related to the other first, second and third neighbor bonds by the appropriate symmetry operations of the crystal.
The Z$_1$ and Z$_3$ bonds are parallel to the crystallographic $a$ direction, in cubic coordinates along $[1\bar{1}0]$ and the corresponding $C_2$ rotation axis is parallel to the bond. The Z$_2$ bond is perpendicular to that bond, in cubic coordinates along the $[11\bar{2}]$ direction. Since in this case the $C_2$ axis is perpendicular to the bond and in-plane, it turns out to be parallel to the $C_2$ axes of the Z$_1$ and Z$_3$ bond and the same restrictions on the hopping matrices are valid in this case. 

In \cref{tab:hoppings} we list the generic hopping matrix on a Z-bond with $2/m$ symmetry at the bond center. 
The hopping matrix between $t_{2g}$ orbitals is fully determined by four parameters $t_{1\ldots4}$, following the convention introduced in Ref.~\onlinecite{rau2014generic}. For a perfect octahedral geometry of the ligands $t_4=0$. In the materials considered, the octahedra do not deviate too strongly from a perfect shape, hence $t_4$ is generally small, as can be seen for the hopping parameters listed for the example case NiCl$_2$. Following from the Slater-Koster integrals in a 90$^\circ$ $M$-$X$-$M$ geometry\cite{pavarini2012}, $t_2$ is dominated by ligand-assisted hopping processes, while $t_1$ and $t_3$ arise mainly from direct hopping. 

Hopping between $t_{2g}$ and $e_g$ orbitals is constrained to three parameters, $\tilde{t}$, $t_7$, $t_8$. 
However, if the considered bond lies within a mirror plane, $t_7$ and $t_8$ vanish. While this is not a symmetry of the crystal, it is a symmetry of a single bond Ni$_2$I$_{10}$ molecule. In the crystal, these hoppings are hence finite, but small. Consequently, the ligand-assisted $\tilde{t}$ dominates the exchange between $t_{2g}$ and $e_g$ orbitals. 

Finally, hopping between $e_g$ orbitals can be described with two parameters $t_5$ and $t_6$, where the off-diagonal terms vanish due to the $2/m$ symmetry. 
These hoppings turn out to be important for further neighbor interactions on e.g. the Z$_3$-bond.

\subsubsection{Effects of spin-orbit coupling}

As mentioned in the main text, the Hamiltonian $\mathcal{H}_{\rm hop}=\sum_{lm} \sum_{\alpha \beta} \underline{c}_{l\alpha}^{\rm T}\ \{ t_{\alpha \beta}^{lm}\ \textrm{I} + \frac{i}{2} (\vec{\lambda}_{\alpha \beta}^{lm} \cdot \vec{\sigma})\}\ \underline{c}_{m \beta}$ is suitable to describe hopping processes including spin-orbit coupling effects. In \cref{tab:hoppings_rel}, we list for the example case VI$_2$ the hopping parameters on-site and for the Z$_1$-bond. 

The on-site hopping parameters can be directly compared to the analytic expressions of the matrix elements in $\lambda_{\rm eff} \mathbf{L} \cdot \mathbf{S}$ for $d$ orbitals in the atomic limit:

\begin{equation}
\begin{array}{r |c c c c c}
\lambda_z						& d_{xy} & d_{xz} & d_{yz} & d_{x^2\text{-}y^2} & d_{z^2}     \\
\hline
d_{xy}				& 0 		& 0      & 0                           & 2 \lambda_{\rm eff}					& 0 	\\			
d_{xz}				& 0 		& 0      & -\lambda_{\rm eff}                          & 0 					& 0 	\\		
d_{yz}				& 0 		& \lambda_{\rm eff}      & 0                           & 0 					& 0 	\\	
d_{x^2\text{-}y^2}		& -2 \lambda_{\rm eff}		& 0      & 0                           & 0 					& 0 	\\		
d_{z^2}				& 0 		& 0      & 0                           & 0 					& 0	\\		
\end{array}
\end{equation}

\begin{equation}
\begin{array}{r |c c c c c}
\lambda_x						& d_{xy} & d_{xz} & d_{yz} & d_{x^2\text{-}y^2} & d_{z^2} \\
\hline
d_{xy}			& 0 	& -\lambda_{\rm eff} & 0 	    & 0		& 0 		 \\ 
d_{xz}			& \lambda_{\rm eff} 	& 0  & 0 	    & 0 		& 0 		\\ 
d_{yz}			& 0 	& 0  & 0 	    & -\lambda_{\rm eff}		& - \sqrt{3}  \lambda_{\rm eff}	\\
d_{x^2\text{-}y^2}	& 0	& 0  & \lambda_{\rm eff} 	    & 0 		& 0 		\\ 
d_{z^2}			& 0 	& 0  &  \sqrt{3} \lambda_{\rm eff} & 0 		& 0		
\end{array}
\end{equation}

\begin{equation}
\begin{array}{r |c c c c c}
\lambda_y	 & d_{xy} & d_{xz} & d_{yz} & d_{x^2\text{-}y^2} & d_{z^2}\\
\hline
d_{xy}					& 0 	& 0          & \lambda_{\rm eff}        & 0		& 0 		\\
d_{xz}					& 0 	& 0          & 0         & -\lambda_{\rm eff}		&  \sqrt{3} \lambda_{\rm eff}	\\
d_{yz}					& -\lambda_{\rm eff} 	& 0          & 0         & 0		& 0 		\\  
d_{x^2\text{-}y^2}			& 0	& \lambda_{\rm eff} 		& 0      & 0		& 0 		\\
\langle d_{z^2} \vert					& 0 	& - \sqrt{3} \lambda_{\rm eff} & 0        & 0		& 0		
\end{array}
\end{equation}
As discussed in the main text, comparison to the values obtained for the monolayer structures (see \cref{tab:hoppings_rel} for the example case VI$_2$) does not allow to identify one unique effective spin-orbit coupling strength $\lambda_{\rm eff}$. This can be attributed to the fact that spin-orbit coupling effects arise mainly from the heavy ligand $p$ elements (here I), not the metal $d$ elements (here V).

On a nearest neighbor bond, one can assume that the crystal-field effects at the ligand site split the p-levels, such that the $p_z$ orbital is split off from the $p_x$ and $p_y$. This suggests that the $z$-component of the SOC at the ligand dominates the effect, which couples the $p_x$ and $p_y$ orbitals. In this case, the SOC at the ligand, when projected into the d-orbital Wannier functions, results in a complex hopping $\lambda^z_{(x^2\text{-}y^2;z^2)}$. A similar analysis of this situation was done in Ref.~\onlinecite{stavropoulos2019microscopic}.

\begin{table}
\centering\def\arraystretch{1.1}
    \begin{ruledtabular}
    \begin{tabular}{R|RRRRR|RRRRR}
    \multicolumn{1}{C}{} & \multicolumn{5}{C}{\text{on-site VI}_2}& \multicolumn{5}{C}{\text{Z}_1\text{-bond VI}_2}  \\
    \lambda_z & d_{xy} & d_{xz} & d_{yz} & d_{x^2\text{-}y^2} & d_{z^2} 
    & d_{xy} & d_{xz} & d_{yz} & d_{x^2\text{-}y^2} & d_{z^2}\\
    \hline
   d_{xy}  		&0&		 0&		 0&		-31&		 0  &	 0&		-1&		 1&		-17&		 0 \\
d_{xz} 			&0&		 0&		\mathbf{-38}&		2&		 0  & 	 1&		 0&		-2&		-1&		-3 \\
d_{yz}                  &0&		\mathbf{38}&		 0&		2&		 0	&-1&		 2&		 0&		-1&		 3 \\
d_{x^2\text{-}y^2}      &31&		-2&		-2&		0&		-6	& 17&		 1&		 1&		 0&		 \mathbf{36}\\
d_{z^2}                 &0&		 0&		 0&		6&		 0	& 0&		 3&		-3&		\mathbf{-36}&		 0 \\
\multicolumn{11}{C}{}  \\[-1.5ex]
\lambda_x & d_{xy} & d_{xz} & d_{yz} & d_{x^2\text{-}y^2} & d_{z^2} 
    & d_{xy} & d_{xz} & d_{yz} & d_{x^2\text{-}y^2} & d_{z^2}\\
    \hline
d_{xy}  		& 0&		\mathbf{-38}&		 0&		-1&		-2&	 0&		-1&		-12&		 1&		-1    \\
d_{xz} 			&\mathbf{38}&		 0&		 0&		 0&		-2&	 1&		 0&		 0&		 21&		-1    \\
d_{yz}                  & 0&		 0&		 0&		16&		27&	 12&		0&		 0&		 0&		-2  \\
d_{x^2\text{-}y^2}      &
1&		 0&		-16&		 0&		-6&	-1&		-21&		0&		 0&		 1   \\
d_{z^2}                 &2&		2&		-27&		6&		 0&	 1&		 1&		 2&		-1&		 0 \\
\multicolumn{11}{C}{}  \\[-1.5ex]
\lambda_y & d_{xy} & d_{xz} & d_{yz} & d_{x^2\text{-}y^2} & d_{z^2} 
    & d_{xy} & d_{xz} & d_{yz} & d_{x^2\text{-}y^2} & d_{z^2}\\
    \hline
d_{xy}  		& 0&		 0&		\mathbf{38}&		-1&		2&	 0&		 12&		 1&		 1&		 1   \\
d_{xz} 			&0&		 0&		 0&		16&		-27&	-12&		 0&		 0&		 0&		 2   \\
d_{yz}                  &\mathbf{-38}&		0&		 0&		-0&		2&	-1&		0&		 0&		 21&		 1    \\
d_{x^2\text{-}y^2}      &1&		-16&		 0&		 0&		-6&	-1&		0&		-21&		 0&		 1   \\
d_{z^2}                 &-2&		27&		-2&		6&		 0&	-1&		-2&		-1&		-1&		 0   \\
    \end{tabular}
    \end{ruledtabular}
    \caption{Relativistic hopping parameters $\lambda_\nu$ (in meV) for VI$_2$ on-site and on a Z$_1$-bond as defined in \cref{fig:structure}, with $\nu \in \{x,y,z\}$  and $\mathcal{H}_{\rm hop} = t_{ij}^{\alpha \beta} + \frac{i}{2} \vec{\lambda}_{ij}^{\alpha \beta} \cdot \vec{\sigma}$.}
    \label{tab:hoppings_rel}
\end{table}

\section{Spin-polarized Wannier function analysis}
\label{sec:spinWan}

\begin{figure}[t!]
    \centering
    \includegraphics[width=0.8\columnwidth]{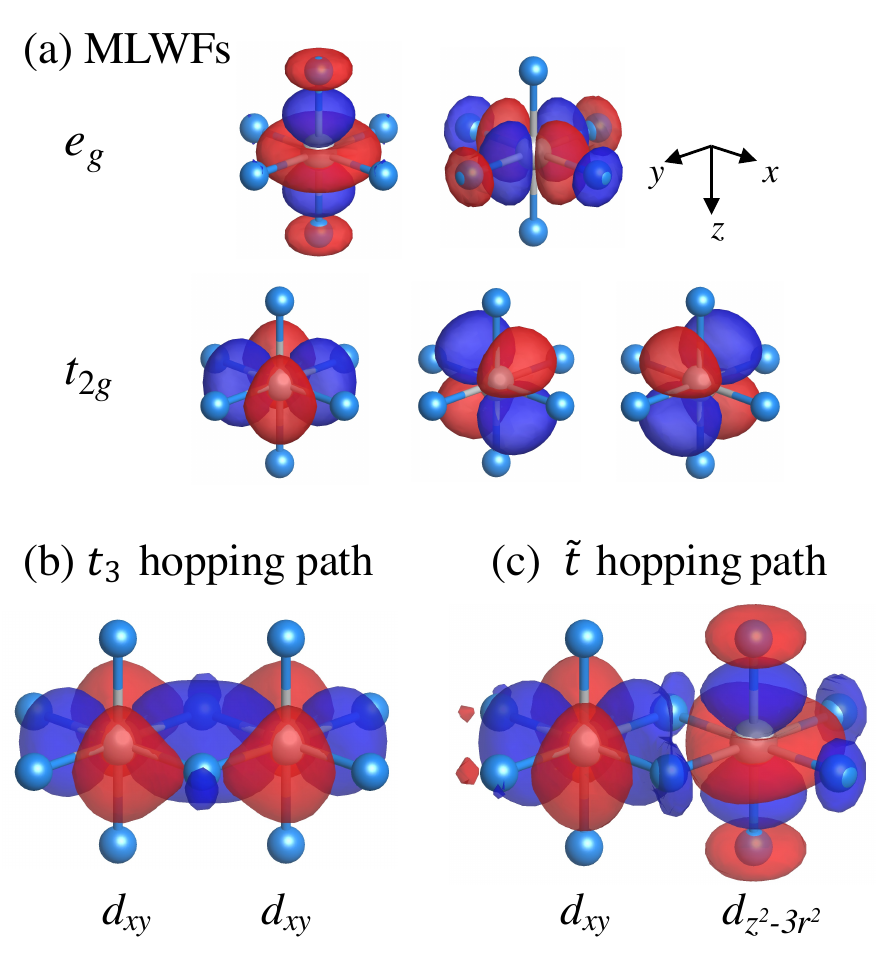}
    \caption{(a) 3$d$-orbital MLWFs in VCl$_6$ octahedral coordination in VCl$_2$ monolayer.  (b, c) Combination of MLWFs for $t_3$ and  $\tilde{t}$ hopping paths in ferromagnetic configuration. Isosurface levels were set at $\pm 0.8$ $a_0^{-3/2}$. Blue and red colors show opposite signs of MLWF. }
    \label{fig:MLWFs} 
\end{figure}

For completion, we discuss in this section spin-polarized nearest neighbor hopping integrals
to give further insight into the microscopic mechanism behind the magnetic couplings in the cases of monolayer $M$Cl$_2$ ($M$=\{V, Mn, Ni\}).
After a DFT calculation was performed by using the VASP code\cite{kresse_paw} and GGA+$U$ functional with $U=1.8$\,eV and $J_{\rm H} = 0.8$\,eV, 
the hopping parameter $t_{dd}^{\rm eff}$ between the transition-metal sites was extracted via the Maximally Localized Wannier Functions (MLWFs) as constructed from projection of five transition-metal $d$ states and six ligand $p$ states~\cite{thao2021}. 
The shapes of the $d$-orbital MLWFs are shown in \cref{fig:MLWFs}(a). Note that the $dp$-$pd$ hybridization process  is implicitly included in $t_{dd}^{\rm eff}$ hopping since the Wannier function has delocalized character reflecting the hybridization with the surrounding ligands’ $p$ orbital states. The orbital-dependent hopping integrals between the first nearest neighbour $M$ sites were calculated in ferromagnetic spin configuration. 

\begin{figure}
    \centering
    \includegraphics[width=0.8\columnwidth]{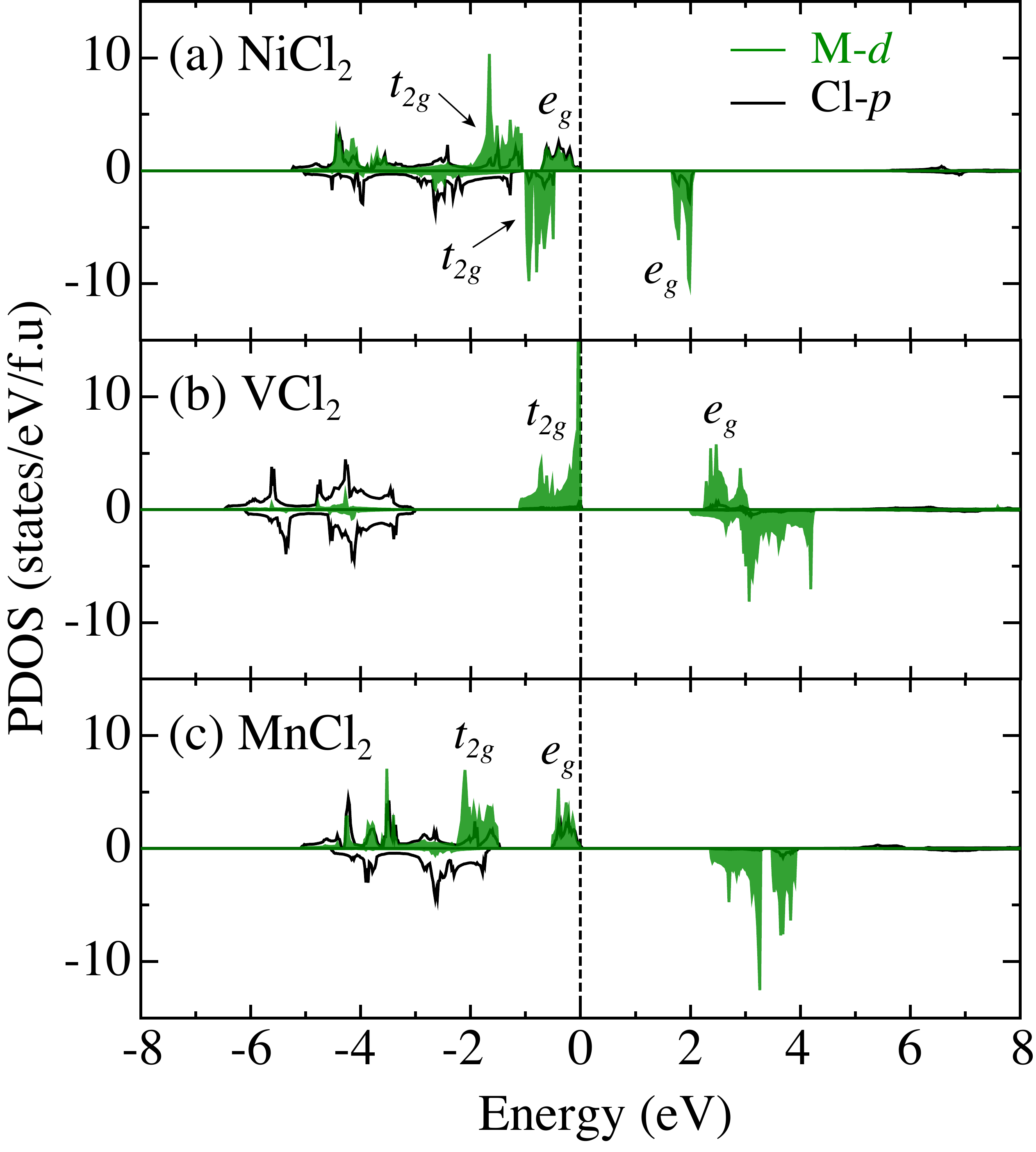}
    \caption{DOS projected onto $M$-$d$ and Cl-$p$ 
     orbital states  for (a) NiCl$_2$, (b) VCl$_2$ and (c) MnCl$_2$. 
     Fermi level is set at energy origin.}
    \label{fig:pdos_pd} 
\end{figure}

Figure \ref{fig:pdos_pd} shows the density of states projected onto $M$-$d$ and Cl-$p$ orbital states.
Strong hybridization between  $M$-$d$ and Cl-$p$ orbitals 
can be seen in the cases of NiCl$_2$ and MnCl$_2$, but not in VCl$_2$, resultant from the different $d$ electron filling. The CFS causes large band gap between $e_g$ and $t_{2g}$ orbital states in Ni minority-spin state and V majority-spin state. 

\begin{table}
    \centering\def\arraystretch{1.1}
    \begin{ruledtabular}
   \begin{tabular}{R|RRR}
    & {\rm NiCl_2} & {\rm VCl_2} & {\rm MnCl_2} \\
    \hline
    t_{\rm 3} & -77 & -250 & -81 \\
    \tilde{t} & 37 & -33 & 0.5\\
    \end{tabular}
    \end{ruledtabular}
    \caption{Hopping integrals $t_{\rm eff}$ (meV) calculated via spin-polarized MLWFs basis set
    for first nearest neighbor coupling in ferromagnetic spin configuration. 
    }
    \label{tab:hopping_compare}
\end{table}

As mentioned in \cref{sec:nearest_exchange}, the nearest neighbor exchange coupling can be explained by the hopping integrals between $d_{xy}$ and $d_{xy}$ orbitals and $d_{xy}$ and $d_{z^2}$ orbitals, namely, $t_3$ and $\tilde{t}$. 
Table~\ref{tab:hopping_compare} lists the calculated hopping parameters through the spin-polarized Wannier functions. 

In NiCl$_2$, weak ferromagnetic exchange is attributed to $\tilde{t}$ (= 37 meV)
hopping between filled $t_{2g}^\downarrow$ state and empty $e_g^\downarrow$ state. 
Since the $t_{2g}$ state is fully occupied, $t_3$ hopping does not play any role in the exchange interaction. 

In VCl$_2$, $t_{2g}^\uparrow$ state is fully occupied while $e_{g}^\uparrow$ state is empty. 
It is found that anti-parallel-spin-favored $t_3$ hopping is much stronger than parallel-spin-favored $\tilde{t}$ hopping. As shown in \cref{fig:MLWFs}(b) and (c), 
$t_3$ hopping is direct $d-d$ hopping while $\tilde{t}$ hopping is $dp(\sigma)$-$pd(\pi)$ indirect hopping. 
This result is consistent with what was discussed in \cref{sec:nearest_exchange}. 

In MnCl$_2$, $d^5$ electron fully occupies the majority spin state and opens a wide band gap. 
As being speculated, this leads to weak exchange interaction while $|t_3| >> |\tilde{t}|$ condition results in anti-parallel spin exchange interaction.

\section{Exchange parameters for alternative $U_{\rm avg}$ and $J_{\rm avg}$ values}
\label{sec:results_Ualt}

To provide context of the dependence on the Coulomb repulsion $U_{\rm avg}$ and Hund's coupling $J_{\rm avg}$, we list in \cref{tab:MX2_Ualt} the exchange parameters extracted with projED and the 4-states method for $(U_{\rm avg},J_{\rm avg})=(1.8,0.8)\,$eV. These parameters were chosen previously by some of the authors\cite{amoroso2020spontaneous,amoroso2021tuning} for the analysis of the Ni$X_2$ materials.

Not surprisingly, the trends agree with the parameters listed in the main text in \cref{tab:NiX2,tab:VX2,tab:MnX2} for the values determined with constrained RPA. The overall tendency toward larger absolute values in both methods can be assigned to the smaller Coulomb repulsion, which results in an overall increase of the magnetic exchange. 

\begin{table}
    \centering\def\arraystretch{1.1}
    \begin{ruledtabular}
    \begin{tabular}{C|CCC|CCC}
      & \multicolumn{3}{c|}{\text{projED}}  & \multicolumn{3}{c}{\text{4-states}} \\
      & \text{NiCl}_2 & \text{NiBr}_2 & \text{NiI}_2  & \text{NiCl}_2 & \text{NiBr}_2 & \text{NiI}_2   \\
    \hline 
    \hline
    J_1 &        -2.2&	-2.1&	-2.5&		-5.1&	-6.0&	-8.0 
    \\
    K_1 &                0.&	+0.1&	+1.3&	0.&	+0.2&	+3.2 \\
    \Gamma_1&             0.& 0.& 0.& 0.& 0.& 0. \\
    \Gamma_1^\prime &     0.& 0.& -0.1& 0.& 0.& -0.2 \\
    J_2 &                 -0.1& -0.1& -0.1& -0.1& -0.1& -0.3 \\
    J_3 &                 +1.2& +1.7&	+2.4&	+1.7& +2.9&	+5.8 \\
    A_c &                 0.&	+0.1&	+0.7&	0.&	0.&	+0.6 \\
    \hline
    \hline
    \multicolumn{7}{C}{}  \\[-1.5ex]
    \hline
    \hline
    & \multicolumn{3}{c|}{\text{projED}}  & \multicolumn{3}{c}{\text{4-states}} \\
    & \text{VCl}_2 & \text{VBr}_2 & \text{VI}_2  & \text{VCl}_2 & \text{VBr}_2 & \text{VI}_2 \\
    \hline
    \hline
    J_1 &+7.8&	+5.0&	+2.6&		+8.6&	+5.1&	+1.8 \\
    K_1 &                 0.& 0.& 0.& 0.& 0.& 0. \\
    \Gamma_1&             0.& 0.& 0.& 0.& 0.& 0. \\
    \Gamma_1^\prime &     0.& 0.& 0.& 0.& 0.& 0. \\
    J_2 &                 0.& +0.1& +0.1& 0.& 0.& +0.1 \\
    J_3 &                 0.& +0.1& +0.1& -0.1& +0.2& +0.3 \\
    A_c &                 0.& 0.& -0.1& 0.& 0. & 0. \\
    \hline
    \hline
    \multicolumn{7}{C}{}  \\[-1.5ex]
    \hline
    \hline
    & \multicolumn{3}{c|}{\text{projED}}  & \multicolumn{3}{c}{\text{4-states}} \\
    & \text{MnCl}_2 & \text{MnBr}_2 & \text{MnI}_2  & \text{MnCl}_2 & \text{MnBr}_2 & \text{MnI}_2 \\
    \hline
    \hline
    J_1 &+1.2&	+1.0&	+0.9&		+0.4&	+0.3&	+0.3 \\
    K_1 &                 0.& 0.& 0.& 0.& 0.& 0. \\
    \Gamma_1&             0.& 0.& 0.& 0.& 0.& 0. \\
    \Gamma_1^\prime &     0.& 0.& 0.& 0.& 0.& 0. \\
    J_2 &                 0.& 0.& +0.1& 0.& 0.& +0.1 \\
    J_3 &                 +0.1& +0.1& +0.1& +0.1& +0.1& +0.1 \\
    A_c &                 0.& 0.& 0.& 0.& 0. & 0.
    \end{tabular}
    \end{ruledtabular}
    \caption{Exchange parameters (in meV) for the $MX_2$ monolayer structures using $(U_{\rm avg},J_{\rm avg})=(1.8,0.8)\,$eV; extracted with the projED method (left) and the 4-states energy mapping method (right).}
    \label{tab:MX2_Ualt}
\end{table}

\bibliography{ref.bib}

\begin{thebibliography}{86}%
\makeatletter
\providecommand \@ifxundefined [1]{%
 \@ifx{#1\undefined}
}%
\providecommand \@ifnum [1]{%
 \ifnum #1\expandafter \@firstoftwo
 \else \expandafter \@secondoftwo
 \fi
}%
\providecommand \@ifx [1]{%
 \ifx #1\expandafter \@firstoftwo
 \else \expandafter \@secondoftwo
 \fi
}%
\providecommand \natexlab [1]{#1}%
\providecommand \enquote  [1]{``#1''}%
\providecommand \bibnamefont  [1]{#1}%
\providecommand \bibfnamefont [1]{#1}%
\providecommand \citenamefont [1]{#1}%
\providecommand \href@noop [0]{\@secondoftwo}%
\providecommand \href [0]{\begingroup \@sanitize@url \@href}%
\providecommand \@href[1]{\@@startlink{#1}\@@href}%
\providecommand \@@href[1]{\endgroup#1\@@endlink}%
\providecommand \@sanitize@url [0]{\catcode `\\12\catcode `\$12\catcode
  `\&12\catcode `\#12\catcode `\^12\catcode `\_12\catcode `\%12\relax}%
\providecommand \@@startlink[1]{}%
\providecommand \@@endlink[0]{}%
\providecommand \url  [0]{\begingroup\@sanitize@url \@url }%
\providecommand \@url [1]{\endgroup\@href {#1}{\urlprefix }}%
\providecommand \urlprefix  [0]{URL }%
\providecommand \Eprint [0]{\href }%
\providecommand \doibase [0]{http://dx.doi.org/}%
\providecommand \selectlanguage [0]{\@gobble}%
\providecommand \bibinfo  [0]{\@secondoftwo}%
\providecommand \bibfield  [0]{\@secondoftwo}%
\providecommand \translation [1]{[#1]}%
\providecommand \BibitemOpen [0]{}%
\providecommand \bibitemStop [0]{}%
\providecommand \bibitemNoStop [0]{.\EOS\space}%
\providecommand \EOS [0]{\spacefactor3000\relax}%
\providecommand \BibitemShut  [1]{\csname bibitem#1\endcsname}%
\let\auto@bib@innerbib\@empty
\bibitem [{\citenamefont {Ramirez}(1994)}]{ramirez1994strongly}%
  \BibitemOpen
  \bibfield  {author} {\bibinfo {author} {\bibfnamefont {A.~P.}\ \bibnamefont
  {Ramirez}},\ }\href {\doibase 10.1146/annurev.ms.24.080194.002321} {\bibfield
   {journal} {\bibinfo  {journal} {Annu. Rev. Mater. Res.}\ }\textbf {\bibinfo
  {volume} {24}},\ \bibinfo {pages} {453} (\bibinfo {year} {1994})}\BibitemShut
  {NoStop}%
\bibitem [{\citenamefont {Balents}(2010)}]{balents2010spin}%
  \BibitemOpen
  \bibfield  {author} {\bibinfo {author} {\bibfnamefont {L.}~\bibnamefont
  {Balents}},\ }\href {\doibase 10.1038/nature08917} {\bibfield  {journal}
  {\bibinfo  {journal} {Nature}\ }\textbf {\bibinfo {volume} {464}},\ \bibinfo
  {pages} {199} (\bibinfo {year} {2010})}\BibitemShut {NoStop}%
\bibitem [{\citenamefont {Lacroix}\ \emph {et~al.}(2011)\citenamefont
  {Lacroix}, \citenamefont {Mendels},\ and\ \citenamefont
  {Mila~(Eds.)}}]{lacroix2011introduction}%
  \BibitemOpen
  \bibfield  {author} {\bibinfo {author} {\bibfnamefont {C.}~\bibnamefont
  {Lacroix}}, \bibinfo {author} {\bibfnamefont {P.}~\bibnamefont {Mendels}}, \
  and\ \bibinfo {author} {\bibfnamefont {F.}~\bibnamefont {Mila~(Eds.)}},\
  }\href@noop {} {\emph {\bibinfo {title} {Introduction to {F}rustrated
  {M}agnetism}}},\ \bibinfo {series} {Springer Series in Solid-State Sciences},
  Vol.\ \bibinfo {volume} {164}\ (\bibinfo {address} {Berlin},\ \bibinfo {year}
  {2011})\BibitemShut {NoStop}%
\bibitem [{\citenamefont {Starykh}(2015)}]{starykh2015unusual}%
  \BibitemOpen
  \bibfield  {author} {\bibinfo {author} {\bibfnamefont {O.~A.}\ \bibnamefont
  {Starykh}},\ }\href {\doibase 10.1088/0034-4885/78/5/052502} {\bibfield
  {journal} {\bibinfo  {journal} {Rep. Prog. Phys.}\ }\textbf {\bibinfo
  {volume} {78}},\ \bibinfo {pages} {052502} (\bibinfo {year}
  {2015})}\BibitemShut {NoStop}%
\bibitem [{\citenamefont {Liu}\ and\ \citenamefont
  {Xu~(Eds.)}(2019)}]{liu2019spintronic}%
  \BibitemOpen
  \bibfield  {author} {\bibinfo {author} {\bibfnamefont {W.}~\bibnamefont
  {Liu}}\ and\ \bibinfo {author} {\bibfnamefont {Y.}~\bibnamefont
  {Xu~(Eds.)}},\ }\href@noop {} {\emph {\bibinfo {title} {Spintronic 2D
  Materials: Fundamentals and Applications}}},\ Materials Today\ (\bibinfo
  {publisher} {Elsevier Science},\ \bibinfo {year} {2019})\BibitemShut
  {NoStop}%
\bibitem [{\citenamefont {Vasiliev}\ \emph {et~al.}(2019)\citenamefont
  {Vasiliev}, \citenamefont {Volkova}, \citenamefont {Zvereva},\ and\
  \citenamefont {Markina~(Eds.)}}]{vasiliev2019low}%
  \BibitemOpen
  \bibfield  {author} {\bibinfo {author} {\bibfnamefont {A.~N.}\ \bibnamefont
  {Vasiliev}}, \bibinfo {author} {\bibfnamefont {O.~S.}\ \bibnamefont
  {Volkova}}, \bibinfo {author} {\bibfnamefont {E.~A.}\ \bibnamefont
  {Zvereva}}, \ and\ \bibinfo {author} {\bibfnamefont {M.~M.}\ \bibnamefont
  {Markina~(Eds.)}},\ }\href@noop {} {\emph {\bibinfo {title} {Low-Dimensional
  Magnetism}}}\ (\bibinfo  {publisher} {CRC Press},\ \bibinfo {year}
  {2019})\BibitemShut {NoStop}%
\bibitem [{\citenamefont {Winter}\ \emph {et~al.}(2016)\citenamefont {Winter},
  \citenamefont {Li}, \citenamefont {Jeschke},\ and\ \citenamefont
  {Valent\'{\i}}}]{winter2016challenges}%
  \BibitemOpen
  \bibfield  {author} {\bibinfo {author} {\bibfnamefont {S.~M.}\ \bibnamefont
  {Winter}}, \bibinfo {author} {\bibfnamefont {Y.}~\bibnamefont {Li}}, \bibinfo
  {author} {\bibfnamefont {H.~O.}\ \bibnamefont {Jeschke}}, \ and\ \bibinfo
  {author} {\bibfnamefont {R.}~\bibnamefont {Valent\'{\i}}},\ }\href {\doibase
  10.1103/PhysRevB.93.214431} {\bibfield  {journal} {\bibinfo  {journal} {Phys.
  Rev. B}\ }\textbf {\bibinfo {volume} {93}},\ \bibinfo {pages} {214431}
  (\bibinfo {year} {2016})}\BibitemShut {NoStop}%
\bibitem [{\citenamefont {Winter}\ \emph
  {et~al.}(2017{\natexlab{a}})\citenamefont {Winter}, \citenamefont {Riedl},
  \citenamefont {Maksimov}, \citenamefont {Chernyshev}, \citenamefont
  {Honecker},\ and\ \citenamefont {Valent\'{\i}}}]{winter2017breakdown}%
  \BibitemOpen
  \bibfield  {author} {\bibinfo {author} {\bibfnamefont {S.~M.}\ \bibnamefont
  {Winter}}, \bibinfo {author} {\bibfnamefont {K.}~\bibnamefont {Riedl}},
  \bibinfo {author} {\bibfnamefont {P.~A.}\ \bibnamefont {Maksimov}}, \bibinfo
  {author} {\bibfnamefont {A.~L.}\ \bibnamefont {Chernyshev}}, \bibinfo
  {author} {\bibfnamefont {A.}~\bibnamefont {Honecker}}, \ and\ \bibinfo
  {author} {\bibfnamefont {R.}~\bibnamefont {Valent\'{\i}}},\ }\href {\doibase
  10.1038/s41467-017-01177-0} {\bibfield  {journal} {\bibinfo  {journal} {Nat.
  Commun.}\ }\textbf {\bibinfo {volume} {8}},\ \bibinfo {pages} {1152}
  (\bibinfo {year} {2017}{\natexlab{a}})}\BibitemShut {NoStop}%
\bibitem [{\citenamefont {Kim}\ and\ \citenamefont
  {Kee}(2016)}]{kim2016crystal}%
  \BibitemOpen
  \bibfield  {author} {\bibinfo {author} {\bibfnamefont {H.-S.}\ \bibnamefont
  {Kim}}\ and\ \bibinfo {author} {\bibfnamefont {H.-Y.}\ \bibnamefont {Kee}},\
  }\href {\doibase 10.1103/PhysRevB.93.155143} {\bibfield  {journal} {\bibinfo
  {journal} {Phys. Rev. B}\ }\textbf {\bibinfo {volume} {93}},\ \bibinfo
  {pages} {155143} (\bibinfo {year} {2016})}\BibitemShut {NoStop}%
\bibitem [{\citenamefont {Yadav}\ \emph {et~al.}(2016)\citenamefont {Yadav},
  \citenamefont {Bogdanov}, \citenamefont {Katukuri}, \citenamefont
  {Nishimoto}, \citenamefont {van~den Brink},\ and\ \citenamefont
  {Hozoi}}]{yadav2016kitaev}%
  \BibitemOpen
  \bibfield  {author} {\bibinfo {author} {\bibfnamefont {R.}~\bibnamefont
  {Yadav}}, \bibinfo {author} {\bibfnamefont {N.~A.}\ \bibnamefont {Bogdanov}},
  \bibinfo {author} {\bibfnamefont {V.~M.}\ \bibnamefont {Katukuri}}, \bibinfo
  {author} {\bibfnamefont {S.}~\bibnamefont {Nishimoto}}, \bibinfo {author}
  {\bibfnamefont {J.}~\bibnamefont {van~den Brink}}, \ and\ \bibinfo {author}
  {\bibfnamefont {L.}~\bibnamefont {Hozoi}},\ }\href {\doibase
  10.1038/srep37925} {\bibfield  {journal} {\bibinfo  {journal} {Sci. Rep.}\
  }\textbf {\bibinfo {volume} {6}},\ \bibinfo {pages} {37925} (\bibinfo {year}
  {2016})}\BibitemShut {NoStop}%
\bibitem [{\citenamefont {Hou}\ \emph {et~al.}(2017)\citenamefont {Hou},
  \citenamefont {Xiang},\ and\ \citenamefont {Gong}}]{hou2017unveiling}%
  \BibitemOpen
  \bibfield  {author} {\bibinfo {author} {\bibfnamefont {Y.~S.}\ \bibnamefont
  {Hou}}, \bibinfo {author} {\bibfnamefont {H.~J.}\ \bibnamefont {Xiang}}, \
  and\ \bibinfo {author} {\bibfnamefont {X.~G.}\ \bibnamefont {Gong}},\ }\href
  {\doibase 10.1103/PhysRevB.96.054410} {\bibfield  {journal} {\bibinfo
  {journal} {Phys. Rev. B}\ }\textbf {\bibinfo {volume} {96}},\ \bibinfo
  {pages} {054410} (\bibinfo {year} {2017})}\BibitemShut {NoStop}%
\bibitem [{\citenamefont {Eichstaedt}\ \emph {et~al.}(2019)\citenamefont
  {Eichstaedt}, \citenamefont {Zhang}, \citenamefont {Laurell}, \citenamefont
  {Okamoto}, \citenamefont {Eguiluz},\ and\ \citenamefont
  {Berlijn}}]{eichstaedt2019deriving}%
  \BibitemOpen
  \bibfield  {author} {\bibinfo {author} {\bibfnamefont {C.}~\bibnamefont
  {Eichstaedt}}, \bibinfo {author} {\bibfnamefont {Y.}~\bibnamefont {Zhang}},
  \bibinfo {author} {\bibfnamefont {P.}~\bibnamefont {Laurell}}, \bibinfo
  {author} {\bibfnamefont {S.}~\bibnamefont {Okamoto}}, \bibinfo {author}
  {\bibfnamefont {A.~G.}\ \bibnamefont {Eguiluz}}, \ and\ \bibinfo {author}
  {\bibfnamefont {T.}~\bibnamefont {Berlijn}},\ }\href {\doibase
  10.1103/PhysRevB.100.075110} {\bibfield  {journal} {\bibinfo  {journal}
  {Phys. Rev. B}\ }\textbf {\bibinfo {volume} {100}},\ \bibinfo {pages}
  {075110} (\bibinfo {year} {2019})}\BibitemShut {NoStop}%
\bibitem [{\citenamefont {Laurell}\ and\ \citenamefont
  {Okamoto}(2020)}]{laurell2020dynamical}%
  \BibitemOpen
  \bibfield  {author} {\bibinfo {author} {\bibfnamefont {P.}~\bibnamefont
  {Laurell}}\ and\ \bibinfo {author} {\bibfnamefont {S.}~\bibnamefont
  {Okamoto}},\ }\href {\doibase 10.1038/s41535-019-0203-y} {\bibfield
  {journal} {\bibinfo  {journal} {npj Quantum Mater.}\ }\textbf {\bibinfo
  {volume} {5}},\ \bibinfo {pages} {2} (\bibinfo {year} {2020})}\BibitemShut
  {NoStop}%
\bibitem [{\citenamefont {Lado}\ and\ \citenamefont
  {Fern{\'{a}}ndez-Rossier}(2017)}]{lado2017origin}%
  \BibitemOpen
  \bibfield  {author} {\bibinfo {author} {\bibfnamefont {J.~L.}\ \bibnamefont
  {Lado}}\ and\ \bibinfo {author} {\bibfnamefont {J.}~\bibnamefont
  {Fern{\'{a}}ndez-Rossier}},\ }\href {\doibase 10.1088/2053-1583/aa75ed}
  {\bibfield  {journal} {\bibinfo  {journal} {2D Mater.}\ }\textbf {\bibinfo
  {volume} {4}},\ \bibinfo {pages} {035002} (\bibinfo {year}
  {2017})}\BibitemShut {NoStop}%
\bibitem [{\citenamefont {Xu}\ \emph {et~al.}(2018)\citenamefont {Xu},
  \citenamefont {Feng}, \citenamefont {Xiang},\ and\ \citenamefont
  {Bellaiche}}]{xu2018interplay}%
  \BibitemOpen
  \bibfield  {author} {\bibinfo {author} {\bibfnamefont {C.}~\bibnamefont
  {Xu}}, \bibinfo {author} {\bibfnamefont {J.}~\bibnamefont {Feng}}, \bibinfo
  {author} {\bibfnamefont {H.}~\bibnamefont {Xiang}}, \ and\ \bibinfo {author}
  {\bibfnamefont {L.}~\bibnamefont {Bellaiche}},\ }\href {\doibase
  10.1038/s41524-018-0115-6} {\bibfield  {journal} {\bibinfo  {journal} {npj
  Computational Materials}\ }\textbf {\bibinfo {volume} {4}},\ \bibinfo {pages}
  {57} (\bibinfo {year} {2018})}\BibitemShut {NoStop}%
\bibitem [{\citenamefont {Besbes}\ \emph {et~al.}(2019)\citenamefont {Besbes},
  \citenamefont {Nikolaev}, \citenamefont {Meskini},\ and\ \citenamefont
  {Solovyev}}]{besbes2019microscopic}%
  \BibitemOpen
  \bibfield  {author} {\bibinfo {author} {\bibfnamefont {O.}~\bibnamefont
  {Besbes}}, \bibinfo {author} {\bibfnamefont {S.}~\bibnamefont {Nikolaev}},
  \bibinfo {author} {\bibfnamefont {N.}~\bibnamefont {Meskini}}, \ and\
  \bibinfo {author} {\bibfnamefont {I.}~\bibnamefont {Solovyev}},\ }\href
  {\doibase 10.1103/PhysRevB.99.104432} {\bibfield  {journal} {\bibinfo
  {journal} {Phys. Rev. B}\ }\textbf {\bibinfo {volume} {99}},\ \bibinfo
  {pages} {104432} (\bibinfo {year} {2019})}\BibitemShut {NoStop}%
\bibitem [{\citenamefont {Lee}\ \emph {et~al.}(2020)\citenamefont {Lee},
  \citenamefont {Utermohlen}, \citenamefont {Weber}, \citenamefont {Hwang},
  \citenamefont {Zhang}, \citenamefont {van Tol}, \citenamefont {Goldberger},
  \citenamefont {Trivedi},\ and\ \citenamefont {Hammel}}]{lee2020fundamental}%
  \BibitemOpen
  \bibfield  {author} {\bibinfo {author} {\bibfnamefont {I.}~\bibnamefont
  {Lee}}, \bibinfo {author} {\bibfnamefont {F.~G.}\ \bibnamefont {Utermohlen}},
  \bibinfo {author} {\bibfnamefont {D.}~\bibnamefont {Weber}}, \bibinfo
  {author} {\bibfnamefont {K.}~\bibnamefont {Hwang}}, \bibinfo {author}
  {\bibfnamefont {C.}~\bibnamefont {Zhang}}, \bibinfo {author} {\bibfnamefont
  {J.}~\bibnamefont {van Tol}}, \bibinfo {author} {\bibfnamefont {J.~E.}\
  \bibnamefont {Goldberger}}, \bibinfo {author} {\bibfnamefont
  {N.}~\bibnamefont {Trivedi}}, \ and\ \bibinfo {author} {\bibfnamefont
  {P.~C.}\ \bibnamefont {Hammel}},\ }\href {\doibase
  10.1103/PhysRevLett.124.017201} {\bibfield  {journal} {\bibinfo  {journal}
  {Phys. Rev. Lett.}\ }\textbf {\bibinfo {volume} {124}},\ \bibinfo {pages}
  {017201} (\bibinfo {year} {2020})}\BibitemShut {NoStop}%
\bibitem [{\citenamefont {Kartsev}\ \emph {et~al.}(2020)\citenamefont
  {Kartsev}, \citenamefont {Augustin}, \citenamefont {Evans}, \citenamefont
  {Novoselov},\ and\ \citenamefont {Santos}}]{kartsev2020biquadratic}%
  \BibitemOpen
  \bibfield  {author} {\bibinfo {author} {\bibfnamefont {A.}~\bibnamefont
  {Kartsev}}, \bibinfo {author} {\bibfnamefont {M.}~\bibnamefont {Augustin}},
  \bibinfo {author} {\bibfnamefont {R.~F.}\ \bibnamefont {Evans}}, \bibinfo
  {author} {\bibfnamefont {K.~S.}\ \bibnamefont {Novoselov}}, \ and\ \bibinfo
  {author} {\bibfnamefont {E.~J.~G.}\ \bibnamefont {Santos}},\ }\href@noop {}
  {\bibfield  {journal} {\bibinfo  {journal} {npj Comput. Mater.}\ }\textbf
  {\bibinfo {volume} {6}},\ \bibinfo {pages} {1} (\bibinfo {year}
  {2020})}\BibitemShut {NoStop}%
\bibitem [{\citenamefont {Kvashnin}\ \emph {et~al.}(2020)\citenamefont
  {Kvashnin}, \citenamefont {Bergman}, \citenamefont {Lichtenstein},\ and\
  \citenamefont {Katsnelson}}]{kvashnin2020relativistic}%
  \BibitemOpen
  \bibfield  {author} {\bibinfo {author} {\bibfnamefont {Y.~O.}\ \bibnamefont
  {Kvashnin}}, \bibinfo {author} {\bibfnamefont {A.}~\bibnamefont {Bergman}},
  \bibinfo {author} {\bibfnamefont {A.~I.}\ \bibnamefont {Lichtenstein}}, \
  and\ \bibinfo {author} {\bibfnamefont {M.~I.}\ \bibnamefont {Katsnelson}},\
  }\href {\doibase 10.1103/PhysRevB.102.115162} {\bibfield  {journal} {\bibinfo
   {journal} {Phys. Rev. B}\ }\textbf {\bibinfo {volume} {102}},\ \bibinfo
  {pages} {115162} (\bibinfo {year} {2020})}\BibitemShut {NoStop}%
\bibitem [{\citenamefont {Stavropoulos}\ \emph {et~al.}(2021)\citenamefont
  {Stavropoulos}, \citenamefont {Liu},\ and\ \citenamefont
  {Kee}}]{stavropoulos2021magnetic}%
  \BibitemOpen
  \bibfield  {author} {\bibinfo {author} {\bibfnamefont {P.~P.}\ \bibnamefont
  {Stavropoulos}}, \bibinfo {author} {\bibfnamefont {X.}~\bibnamefont {Liu}}, \
  and\ \bibinfo {author} {\bibfnamefont {H.-Y.}\ \bibnamefont {Kee}},\ }\href
  {\doibase 10.1103/PhysRevResearch.3.013216} {\bibfield  {journal} {\bibinfo
  {journal} {Phys. Rev. Research}\ }\textbf {\bibinfo {volume} {3}},\ \bibinfo
  {pages} {013216} (\bibinfo {year} {2021})}\BibitemShut {NoStop}%
\bibitem [{\citenamefont {Edstr\"om}\ \emph {et~al.}(2022)\citenamefont
  {Edstr\"om}, \citenamefont {Amoroso}, \citenamefont {Picozzi}, \citenamefont
  {Barone},\ and\ \citenamefont {Stengel}}]{edstrom2021curved}%
  \BibitemOpen
  \bibfield  {author} {\bibinfo {author} {\bibfnamefont {A.}~\bibnamefont
  {Edstr\"om}}, \bibinfo {author} {\bibfnamefont {D.}~\bibnamefont {Amoroso}},
  \bibinfo {author} {\bibfnamefont {S.}~\bibnamefont {Picozzi}}, \bibinfo
  {author} {\bibfnamefont {P.}~\bibnamefont {Barone}}, \ and\ \bibinfo {author}
  {\bibfnamefont {M.}~\bibnamefont {Stengel}},\ }\href {\doibase
  10.1103/PhysRevLett.128.177202} {\bibfield  {journal} {\bibinfo  {journal}
  {Phys. Rev. Lett.}\ }\textbf {\bibinfo {volume} {128}},\ \bibinfo {pages}
  {177202} (\bibinfo {year} {2022})}\BibitemShut {NoStop}%
\bibitem [{\citenamefont {Tokunaga}\ \emph {et~al.}(2011)\citenamefont
  {Tokunaga}, \citenamefont {Okuyama}, \citenamefont {Kurumaji}, \citenamefont
  {Arima}, \citenamefont {Nakao}, \citenamefont {Murakami}, \citenamefont
  {Taguchi},\ and\ \citenamefont {Tokura}}]{tokunaga2011multiferroicity}%
  \BibitemOpen
  \bibfield  {author} {\bibinfo {author} {\bibfnamefont {Y.}~\bibnamefont
  {Tokunaga}}, \bibinfo {author} {\bibfnamefont {D.}~\bibnamefont {Okuyama}},
  \bibinfo {author} {\bibfnamefont {T.}~\bibnamefont {Kurumaji}}, \bibinfo
  {author} {\bibfnamefont {T.}~\bibnamefont {Arima}}, \bibinfo {author}
  {\bibfnamefont {H.}~\bibnamefont {Nakao}}, \bibinfo {author} {\bibfnamefont
  {Y.}~\bibnamefont {Murakami}}, \bibinfo {author} {\bibfnamefont
  {Y.}~\bibnamefont {Taguchi}}, \ and\ \bibinfo {author} {\bibfnamefont
  {Y.}~\bibnamefont {Tokura}},\ }\href {\doibase 10.1103/PhysRevB.84.060406}
  {\bibfield  {journal} {\bibinfo  {journal} {Phys. Rev. B}\ }\textbf {\bibinfo
  {volume} {84}},\ \bibinfo {pages} {060406} (\bibinfo {year}
  {2011})}\BibitemShut {NoStop}%
\bibitem [{\citenamefont {Kurumaji}\ \emph {et~al.}(2013)\citenamefont
  {Kurumaji}, \citenamefont {Seki}, \citenamefont {Ishiwata}, \citenamefont
  {Murakawa}, \citenamefont {Kaneko},\ and\ \citenamefont
  {Tokura}}]{kurumaji2013magnetoelectric}%
  \BibitemOpen
  \bibfield  {author} {\bibinfo {author} {\bibfnamefont {T.}~\bibnamefont
  {Kurumaji}}, \bibinfo {author} {\bibfnamefont {S.}~\bibnamefont {Seki}},
  \bibinfo {author} {\bibfnamefont {S.}~\bibnamefont {Ishiwata}}, \bibinfo
  {author} {\bibfnamefont {H.}~\bibnamefont {Murakawa}}, \bibinfo {author}
  {\bibfnamefont {Y.}~\bibnamefont {Kaneko}}, \ and\ \bibinfo {author}
  {\bibfnamefont {Y.}~\bibnamefont {Tokura}},\ }\href {\doibase
  10.1103/PhysRevB.87.014429} {\bibfield  {journal} {\bibinfo  {journal} {Phys.
  Rev. B}\ }\textbf {\bibinfo {volume} {87}},\ \bibinfo {pages} {014429}
  (\bibinfo {year} {2013})}\BibitemShut {NoStop}%
\bibitem [{\citenamefont {Kurumaji}\ \emph {et~al.}(2011)\citenamefont
  {Kurumaji}, \citenamefont {Seki}, \citenamefont {Ishiwata}, \citenamefont
  {Murakawa}, \citenamefont {Tokunaga}, \citenamefont {Kaneko},\ and\
  \citenamefont {Tokura}}]{kurumaji2011magnetic}%
  \BibitemOpen
  \bibfield  {author} {\bibinfo {author} {\bibfnamefont {T.}~\bibnamefont
  {Kurumaji}}, \bibinfo {author} {\bibfnamefont {S.}~\bibnamefont {Seki}},
  \bibinfo {author} {\bibfnamefont {S.}~\bibnamefont {Ishiwata}}, \bibinfo
  {author} {\bibfnamefont {H.}~\bibnamefont {Murakawa}}, \bibinfo {author}
  {\bibfnamefont {Y.}~\bibnamefont {Tokunaga}}, \bibinfo {author}
  {\bibfnamefont {Y.}~\bibnamefont {Kaneko}}, \ and\ \bibinfo {author}
  {\bibfnamefont {Y.}~\bibnamefont {Tokura}},\ }\href {\doibase
  10.1103/PhysRevLett.106.167206} {\bibfield  {journal} {\bibinfo  {journal}
  {Phys. Rev. Lett.}\ }\textbf {\bibinfo {volume} {106}},\ \bibinfo {pages}
  {167206} (\bibinfo {year} {2011})}\BibitemShut {NoStop}%
\bibitem [{\citenamefont {Li}\ \emph {et~al.}(2020)\citenamefont {Li},
  \citenamefont {Chen}, \citenamefont {Dong}, \citenamefont {Qiao},
  \citenamefont {He}, \citenamefont {Xiong}, \citenamefont {Li}, \citenamefont
  {Peng}, \citenamefont {Zheng}, \citenamefont {Wang} \emph
  {et~al.}}]{li2020magnetic}%
  \BibitemOpen
  \bibfield  {author} {\bibinfo {author} {\bibfnamefont {Y.}~\bibnamefont
  {Li}}, \bibinfo {author} {\bibfnamefont {D.}~\bibnamefont {Chen}}, \bibinfo
  {author} {\bibfnamefont {X.}~\bibnamefont {Dong}}, \bibinfo {author}
  {\bibfnamefont {L.}~\bibnamefont {Qiao}}, \bibinfo {author} {\bibfnamefont
  {Y.}~\bibnamefont {He}}, \bibinfo {author} {\bibfnamefont {X.}~\bibnamefont
  {Xiong}}, \bibinfo {author} {\bibfnamefont {J.}~\bibnamefont {Li}}, \bibinfo
  {author} {\bibfnamefont {X.}~\bibnamefont {Peng}}, \bibinfo {author}
  {\bibfnamefont {J.}~\bibnamefont {Zheng}}, \bibinfo {author} {\bibfnamefont
  {X.}~\bibnamefont {Wang}},  \emph {et~al.},\ }\href@noop {} {\bibfield
  {journal} {\bibinfo  {journal} {J. Phys. Condens. Matter}\ }\textbf {\bibinfo
  {volume} {32}},\ \bibinfo {pages} {335803} (\bibinfo {year}
  {2020})}\BibitemShut {NoStop}%
\bibitem [{\citenamefont {Song}\ \emph {et~al.}(2022)\citenamefont {Song},
  \citenamefont {Occhialini}, \citenamefont {Ergeçen}, \citenamefont {Ilyas},
  \citenamefont {Amoroso}, \citenamefont {Barone}, \citenamefont {Kapeghian},
  \citenamefont {Watanabe}, \citenamefont {Taniguchi}, \citenamefont {Takashi},
  \citenamefont {Botana}, \citenamefont {Picozzi}, \citenamefont {Gedik},\ and\
  \citenamefont {Comin}}]{song2022experimental}%
  \BibitemOpen
  \bibfield  {author} {\bibinfo {author} {\bibfnamefont {Q.}~\bibnamefont
  {Song}}, \bibinfo {author} {\bibfnamefont {C.~A.}\ \bibnamefont
  {Occhialini}}, \bibinfo {author} {\bibfnamefont {E.}~\bibnamefont
  {Ergeçen}}, \bibinfo {author} {\bibfnamefont {B.}~\bibnamefont {Ilyas}},
  \bibinfo {author} {\bibfnamefont {D.}~\bibnamefont {Amoroso}}, \bibinfo
  {author} {\bibfnamefont {P.}~\bibnamefont {Barone}}, \bibinfo {author}
  {\bibfnamefont {J.}~\bibnamefont {Kapeghian}}, \bibinfo {author}
  {\bibfnamefont {K.}~\bibnamefont {Watanabe}}, \bibinfo {author} {\bibnamefont
  {Taniguchi}}, \bibinfo {author} {\bibnamefont {Takashi}}, \bibinfo {author}
  {\bibfnamefont {A.~S.}\ \bibnamefont {Botana}}, \bibinfo {author}
  {\bibfnamefont {S.}~\bibnamefont {Picozzi}}, \bibinfo {author} {\bibfnamefont
  {N.}~\bibnamefont {Gedik}}, \ and\ \bibinfo {author} {\bibfnamefont
  {R.}~\bibnamefont {Comin}},\ }\href {\doibase 10.1038/s41586-021-04337-x}
  {\bibfield  {journal} {\bibinfo  {journal} {Nature}\ }\textbf {\bibinfo
  {volume} {602}},\ \bibinfo {pages} {601–605} (\bibinfo {year}
  {2022)})}\BibitemShut {NoStop}%
\bibitem [{\citenamefont {Fumega}\ and\ \citenamefont
  {Lado}(2022)}]{Fumega_2022}%
  \BibitemOpen
  \bibfield  {author} {\bibinfo {author} {\bibfnamefont {A.~O.}\ \bibnamefont
  {Fumega}}\ and\ \bibinfo {author} {\bibfnamefont {J.~L.}\ \bibnamefont
  {Lado}},\ }\href {\doibase 10.1088/2053-1583/ac4e9d} {\bibfield  {journal}
  {\bibinfo  {journal} {2D Mater.}\ }\textbf {\bibinfo {volume} {9}},\ \bibinfo
  {pages} {025010} (\bibinfo {year} {2022})}\BibitemShut {NoStop}%
\bibitem [{\citenamefont {Jiang}\ \emph {et~al.}(2019)\citenamefont {Jiang},
  \citenamefont {Li}, \citenamefont {Duan},\ and\ \citenamefont
  {Zhang}}]{jiang2019half}%
  \BibitemOpen
  \bibfield  {author} {\bibinfo {author} {\bibfnamefont {Z.}~\bibnamefont
  {Jiang}}, \bibinfo {author} {\bibfnamefont {Y.}~\bibnamefont {Li}}, \bibinfo
  {author} {\bibfnamefont {W.}~\bibnamefont {Duan}}, \ and\ \bibinfo {author}
  {\bibfnamefont {S.}~\bibnamefont {Zhang}},\ }\href {\doibase
  10.1103/PhysRevLett.122.236402} {\bibfield  {journal} {\bibinfo  {journal}
  {Phys. Rev. Lett.}\ }\textbf {\bibinfo {volume} {122}},\ \bibinfo {pages}
  {236402} (\bibinfo {year} {2019})}\BibitemShut {NoStop}%
\bibitem [{\citenamefont {Botana}\ and\ \citenamefont
  {Norman}(2019)}]{botana2019electronic}%
  \BibitemOpen
  \bibfield  {author} {\bibinfo {author} {\bibfnamefont {A.~S.}\ \bibnamefont
  {Botana}}\ and\ \bibinfo {author} {\bibfnamefont {M.~R.}\ \bibnamefont
  {Norman}},\ }\href {\doibase 10.1103/PhysRevMaterials.3.044001} {\bibfield
  {journal} {\bibinfo  {journal} {Phys. Rev. Materials}\ }\textbf {\bibinfo
  {volume} {3}},\ \bibinfo {pages} {044001} (\bibinfo {year}
  {2019})}\BibitemShut {NoStop}%
\bibitem [{\citenamefont {Stavropoulos}\ \emph {et~al.}(2019)\citenamefont
  {Stavropoulos}, \citenamefont {Pereira},\ and\ \citenamefont
  {Kee}}]{stavropoulos2019microscopic}%
  \BibitemOpen
  \bibfield  {author} {\bibinfo {author} {\bibfnamefont {P.~P.}\ \bibnamefont
  {Stavropoulos}}, \bibinfo {author} {\bibfnamefont {D.}~\bibnamefont
  {Pereira}}, \ and\ \bibinfo {author} {\bibfnamefont {H.-Y.}\ \bibnamefont
  {Kee}},\ }\href {\doibase 10.1103/PhysRevLett.123.037203} {\bibfield
  {journal} {\bibinfo  {journal} {Phys. Rev. Lett.}\ }\textbf {\bibinfo
  {volume} {123}},\ \bibinfo {pages} {037203} (\bibinfo {year}
  {2019})}\BibitemShut {NoStop}%
\bibitem [{\citenamefont {Riedl}\ \emph {et~al.}(2019)\citenamefont {Riedl},
  \citenamefont {Li}, \citenamefont {Valent{\'i}},\ and\ \citenamefont
  {Winter}}]{riedl2019abinitio}%
  \BibitemOpen
  \bibfield  {author} {\bibinfo {author} {\bibfnamefont {K.}~\bibnamefont
  {Riedl}}, \bibinfo {author} {\bibfnamefont {Y.}~\bibnamefont {Li}}, \bibinfo
  {author} {\bibfnamefont {R.}~\bibnamefont {Valent{\'i}}}, \ and\ \bibinfo
  {author} {\bibfnamefont {S.~M.}\ \bibnamefont {Winter}},\ }\href {\doibase
  https://doi.org/10.1002/pssb.201800684} {\bibfield  {journal} {\bibinfo
  {journal} {Phys. Status Solidi B}\ }\textbf {\bibinfo {volume} {256}},\
  \bibinfo {pages} {1800684} (\bibinfo {year} {2019})}\BibitemShut {NoStop}%
\bibitem [{\citenamefont {Xiang}\ \emph {et~al.}(2013)\citenamefont {Xiang},
  \citenamefont {Lee}, \citenamefont {Koo}, \citenamefont {Gong},\ and\
  \citenamefont {Whangbo}}]{xiang2013magnetic}%
  \BibitemOpen
  \bibfield  {author} {\bibinfo {author} {\bibfnamefont {H.}~\bibnamefont
  {Xiang}}, \bibinfo {author} {\bibfnamefont {C.}~\bibnamefont {Lee}}, \bibinfo
  {author} {\bibfnamefont {H.-J.}\ \bibnamefont {Koo}}, \bibinfo {author}
  {\bibfnamefont {X.}~\bibnamefont {Gong}}, \ and\ \bibinfo {author}
  {\bibfnamefont {M.-H.}\ \bibnamefont {Whangbo}},\ }\href@noop {} {\bibfield
  {journal} {\bibinfo  {journal} {Dalton Trans.}\ }\textbf {\bibinfo {volume}
  {42}},\ \bibinfo {pages} {823} (\bibinfo {year} {2013})}\BibitemShut
  {NoStop}%
\bibitem [{\citenamefont {Li}\ \emph {et~al.}(2021)\citenamefont {Li},
  \citenamefont {Yu}, \citenamefont {Lou}, \citenamefont {Feng}, \citenamefont
  {Whangbo},\ and\ \citenamefont {Xiang}}]{li2021spin}%
  \BibitemOpen
  \bibfield  {author} {\bibinfo {author} {\bibfnamefont {X.}~\bibnamefont
  {Li}}, \bibinfo {author} {\bibfnamefont {H.}~\bibnamefont {Yu}}, \bibinfo
  {author} {\bibfnamefont {F.}~\bibnamefont {Lou}}, \bibinfo {author}
  {\bibfnamefont {J.}~\bibnamefont {Feng}}, \bibinfo {author} {\bibfnamefont
  {M.-H.}\ \bibnamefont {Whangbo}}, \ and\ \bibinfo {author} {\bibfnamefont
  {H.}~\bibnamefont {Xiang}},\ }\href {\doibase 10.3390/molecules26040803}
  {\bibfield  {journal} {\bibinfo  {journal} {Molecules}\ }\textbf {\bibinfo
  {volume} {26}},\ \bibinfo {pages} {803} (\bibinfo {year} {2021})}\BibitemShut
  {NoStop}%
\bibitem [{\citenamefont {Anderson}(1950)}]{Anderson1950}%
  \BibitemOpen
  \bibfield  {author} {\bibinfo {author} {\bibfnamefont {P.~W.}\ \bibnamefont
  {Anderson}},\ }\href {\doibase 10.1103/PhysRev.79.350} {\bibfield  {journal}
  {\bibinfo  {journal} {Phys. Rev.}\ }\textbf {\bibinfo {volume} {79}},\
  \bibinfo {pages} {350} (\bibinfo {year} {1950})}\BibitemShut {NoStop}%
\bibitem [{\citenamefont {Kanamori}(1959)}]{Kanamori1958}%
  \BibitemOpen
  \bibfield  {author} {\bibinfo {author} {\bibfnamefont {J.}~\bibnamefont
  {Kanamori}},\ }\href {\doibase https://doi.org/10.1016/0022-3697(59)90061-7}
  {\bibfield  {journal} {\bibinfo  {journal} {J. Phys. Chem. Solids}\ }\textbf
  {\bibinfo {volume} {10}},\ \bibinfo {pages} {87} (\bibinfo {year}
  {1959})}\BibitemShut {NoStop}%
\bibitem [{\citenamefont {Kitaev}(2006)}]{kitaev2006anyons}%
  \BibitemOpen
  \bibfield  {author} {\bibinfo {author} {\bibfnamefont {A.}~\bibnamefont
  {Kitaev}},\ }\href@noop {} {\bibfield  {journal} {\bibinfo  {journal} {Ann.
  Phys.}\ }\textbf {\bibinfo {volume} {321}},\ \bibinfo {pages} {2} (\bibinfo
  {year} {2006})}\BibitemShut {NoStop}%
\bibitem [{\citenamefont {Winter}\ \emph
  {et~al.}(2017{\natexlab{b}})\citenamefont {Winter}, \citenamefont {Tsirlin},
  \citenamefont {Daghofer}, \citenamefont {van~den Brink}, \citenamefont
  {Singh}, \citenamefont {Gegenwart},\ and\ \citenamefont
  {Valent{\'\i}}}]{winter2017models}%
  \BibitemOpen
  \bibfield  {author} {\bibinfo {author} {\bibfnamefont {S.~M.}\ \bibnamefont
  {Winter}}, \bibinfo {author} {\bibfnamefont {A.~A.}\ \bibnamefont {Tsirlin}},
  \bibinfo {author} {\bibfnamefont {M.}~\bibnamefont {Daghofer}}, \bibinfo
  {author} {\bibfnamefont {J.}~\bibnamefont {van~den Brink}}, \bibinfo {author}
  {\bibfnamefont {Y.}~\bibnamefont {Singh}}, \bibinfo {author} {\bibfnamefont
  {P.}~\bibnamefont {Gegenwart}}, \ and\ \bibinfo {author} {\bibfnamefont
  {R.}~\bibnamefont {Valent{\'\i}}},\ }\href@noop {} {\bibfield  {journal}
  {\bibinfo  {journal} {J. Phys. Condens. Matter}\ }\textbf {\bibinfo {volume}
  {29}},\ \bibinfo {pages} {493002} (\bibinfo {year}
  {2017}{\natexlab{b}})}\BibitemShut {NoStop}%
\bibitem [{\citenamefont {Perdew}\ \emph {et~al.}(1996)\citenamefont {Perdew},
  \citenamefont {Burke},\ and\ \citenamefont {Ernzerhof}}]{pbe}%
  \BibitemOpen
  \bibfield  {author} {\bibinfo {author} {\bibfnamefont {J.~P.}\ \bibnamefont
  {Perdew}}, \bibinfo {author} {\bibfnamefont {K.}~\bibnamefont {Burke}}, \
  and\ \bibinfo {author} {\bibfnamefont {M.}~\bibnamefont {Ernzerhof}},\ }\href
  {\doibase 10.1103/PhysRevLett.77.3865} {\bibfield  {journal} {\bibinfo
  {journal} {Phys. Rev. Lett.}\ }\textbf {\bibinfo {volume} {77}},\ \bibinfo
  {pages} {3865} (\bibinfo {year} {1996})}\BibitemShut {NoStop}%
\bibitem [{\citenamefont {Aryasetiawan}\ \emph {et~al.}(2004)\citenamefont
  {Aryasetiawan}, \citenamefont {Imada}, \citenamefont {Georges}, \citenamefont
  {Kotliar}, \citenamefont {Biermann},\ and\ \citenamefont
  {Lichtenstein}}]{Aryasetiawan2004}%
  \BibitemOpen
  \bibfield  {author} {\bibinfo {author} {\bibfnamefont {F.}~\bibnamefont
  {Aryasetiawan}}, \bibinfo {author} {\bibfnamefont {M.}~\bibnamefont {Imada}},
  \bibinfo {author} {\bibfnamefont {A.}~\bibnamefont {Georges}}, \bibinfo
  {author} {\bibfnamefont {G.}~\bibnamefont {Kotliar}}, \bibinfo {author}
  {\bibfnamefont {S.}~\bibnamefont {Biermann}}, \ and\ \bibinfo {author}
  {\bibfnamefont {A.~I.}\ \bibnamefont {Lichtenstein}},\ }\href {\doibase
  10.1103/PhysRevB.70.195104} {\bibfield  {journal} {\bibinfo  {journal} {Phys.
  Rev. B}\ }\textbf {\bibinfo {volume} {70}},\ \bibinfo {pages} {195104}
  (\bibinfo {year} {2004})}\BibitemShut {NoStop}%
\bibitem [{\citenamefont {Aryasetiawan}\ \emph {et~al.}(2006)\citenamefont
  {Aryasetiawan}, \citenamefont {Karlsson}, \citenamefont {Jepsen},\ and\
  \citenamefont {Sch\"onberger}}]{Aryasetiawan2006}%
  \BibitemOpen
  \bibfield  {author} {\bibinfo {author} {\bibfnamefont {F.}~\bibnamefont
  {Aryasetiawan}}, \bibinfo {author} {\bibfnamefont {K.}~\bibnamefont
  {Karlsson}}, \bibinfo {author} {\bibfnamefont {O.}~\bibnamefont {Jepsen}}, \
  and\ \bibinfo {author} {\bibfnamefont {U.}~\bibnamefont {Sch\"onberger}},\
  }\href {\doibase 10.1103/PhysRevB.74.125106} {\bibfield  {journal} {\bibinfo
  {journal} {Phys. Rev. B}\ }\textbf {\bibinfo {volume} {74}},\ \bibinfo
  {pages} {125106} (\bibinfo {year} {2006})}\BibitemShut {NoStop}%
\bibitem [{\citenamefont {Pavarini}\ \emph {et~al.}(2012)\citenamefont
  {Pavarini}, \citenamefont {Koch}, \citenamefont {Anders},\ and\ \citenamefont
  {Jarrell~(Eds.)}}]{pavarini2012}%
  \BibitemOpen
  \bibfield  {author} {\bibinfo {author} {\bibfnamefont {E.}~\bibnamefont
  {Pavarini}}, \bibinfo {author} {\bibfnamefont {E.}~\bibnamefont {Koch}},
  \bibinfo {author} {\bibfnamefont {F.}~\bibnamefont {Anders}}, \ and\ \bibinfo
  {author} {\bibfnamefont {M.}~\bibnamefont {Jarrell~(Eds.)}},\ }\href
  {http://juser.fz-juelich.de/record/136393} {\emph {\bibinfo {title}
  {Correlated {E}lectrons: {F}rom {M}odels to {M}aterials}}},\ \bibinfo
  {series} {Schriften des Forschungszentrums J\"ulich: Modeling and
  Simulation}, Vol.~\bibinfo {volume} {2}\ (\bibinfo {address} {J\"ulich},\
  \bibinfo {year} {2012})\BibitemShut {NoStop}%
\bibitem [{\citenamefont {Eschrig}\ and\ \citenamefont
  {Koepernik}(2009)}]{eschrig2009tight}%
  \BibitemOpen
  \bibfield  {author} {\bibinfo {author} {\bibfnamefont {H.}~\bibnamefont
  {Eschrig}}\ and\ \bibinfo {author} {\bibfnamefont {K.}~\bibnamefont
  {Koepernik}},\ }\href {\doibase 10.1103/PhysRevB.80.104503} {\bibfield
  {journal} {\bibinfo  {journal} {Phys. Rev. B}\ }\textbf {\bibinfo {volume}
  {80}},\ \bibinfo {pages} {104503} (\bibinfo {year} {2009})}\BibitemShut
  {NoStop}%
\bibitem [{\citenamefont {Koepernik}\ and\ \citenamefont
  {Eschrig}(1999)}]{fplo}%
  \BibitemOpen
  \bibfield  {author} {\bibinfo {author} {\bibfnamefont {K.}~\bibnamefont
  {Koepernik}}\ and\ \bibinfo {author} {\bibfnamefont {H.}~\bibnamefont
  {Eschrig}},\ }\href {\doibase 10.1103/PhysRevB.59.1743} {\bibfield  {journal}
  {\bibinfo  {journal} {Phys. Rev. B}\ }\textbf {\bibinfo {volume} {59}},\
  \bibinfo {pages} {1743} (\bibinfo {year} {1999})}\BibitemShut {NoStop}%
\bibitem [{\citenamefont {Georges}\ \emph {et~al.}(2013)\citenamefont
  {Georges}, \citenamefont {Medici},\ and\ \citenamefont
  {Mravlje}}]{georges2013strong}%
  \BibitemOpen
  \bibfield  {author} {\bibinfo {author} {\bibfnamefont {A.}~\bibnamefont
  {Georges}}, \bibinfo {author} {\bibfnamefont {L.~d.}\ \bibnamefont {Medici}},
  \ and\ \bibinfo {author} {\bibfnamefont {J.}~\bibnamefont {Mravlje}},\ }\href
  {\doibase 10.1146/annurev-conmatphys-020911-125045} {\bibfield  {journal}
  {\bibinfo  {journal} {Annu. Rev. Condens. Matter Phys.}\ }\textbf {\bibinfo
  {volume} {4}},\ \bibinfo {pages} {137} (\bibinfo {year} {2013})}\BibitemShut
  {NoStop}%
\bibitem [{\citenamefont {Abragam}\ and\ \citenamefont
  {Bleaney}(2012)}]{abragam2012electron}%
  \BibitemOpen
  \bibfield  {author} {\bibinfo {author} {\bibfnamefont {A.}~\bibnamefont
  {Abragam}}\ and\ \bibinfo {author} {\bibfnamefont {B.}~\bibnamefont
  {Bleaney}},\ }\href@noop {} {\emph {\bibinfo {title} {Electron paramagnetic
  resonance of transition ions}}}\ (\bibinfo  {publisher} {OUP Oxford},\
  \bibinfo {year} {2012})\BibitemShut {NoStop}%
\bibitem [{\citenamefont {Winter}(2022)}]{winter2022magnetic}%
  \BibitemOpen
  \bibfield  {author} {\bibinfo {author} {\bibfnamefont {S.~M.}\ \bibnamefont
  {Winter}},\ }\href@noop {} {\bibfield  {journal} {\bibinfo  {journal} {arXiv
  preprint arXiv:2204.09856}\ } (\bibinfo {year} {2022})}\BibitemShut {NoStop}%
\bibitem [{\citenamefont {Amoroso}\ \emph {et~al.}(2020)\citenamefont
  {Amoroso}, \citenamefont {Barone},\ and\ \citenamefont
  {Picozzi}}]{amoroso2020spontaneous}%
  \BibitemOpen
  \bibfield  {author} {\bibinfo {author} {\bibfnamefont {D.}~\bibnamefont
  {Amoroso}}, \bibinfo {author} {\bibfnamefont {P.}~\bibnamefont {Barone}}, \
  and\ \bibinfo {author} {\bibfnamefont {S.}~\bibnamefont {Picozzi}},\ }\href
  {\doibase 10.1038/s41467-020-19535-w} {\bibfield  {journal} {\bibinfo
  {journal} {Nat. Commun.}\ }\textbf {\bibinfo {volume} {11}},\ \bibinfo
  {pages} {5784} (\bibinfo {year} {2020})}\BibitemShut {NoStop}%
\bibitem [{\citenamefont {Lindgard}\ \emph {et~al.}(1975)\citenamefont
  {Lindgard}, \citenamefont {Birgeneau}, \citenamefont {Guggenheim},\ and\
  \citenamefont {Als-Nielsen}}]{lindgard1975spin}%
  \BibitemOpen
  \bibfield  {author} {\bibinfo {author} {\bibfnamefont {P.~A.}\ \bibnamefont
  {Lindgard}}, \bibinfo {author} {\bibfnamefont {R.~J.}\ \bibnamefont
  {Birgeneau}}, \bibinfo {author} {\bibfnamefont {H.~J.}\ \bibnamefont
  {Guggenheim}}, \ and\ \bibinfo {author} {\bibfnamefont {J.}~\bibnamefont
  {Als-Nielsen}},\ }\href@noop {} {\bibfield  {journal} {\bibinfo  {journal}
  {J. Phys. C: Solid State Phys.}\ }\textbf {\bibinfo {volume} {8}},\ \bibinfo
  {pages} {1059} (\bibinfo {year} {1975})}\BibitemShut {NoStop}%
\bibitem [{\citenamefont {Day}\ \emph {et~al.}(1976)\citenamefont {Day},
  \citenamefont {Dinsdale}, \citenamefont {Krausz},\ and\ \citenamefont
  {Robbins}}]{day1976optical}%
  \BibitemOpen
  \bibfield  {author} {\bibinfo {author} {\bibfnamefont {P.}~\bibnamefont
  {Day}}, \bibinfo {author} {\bibfnamefont {A.}~\bibnamefont {Dinsdale}},
  \bibinfo {author} {\bibfnamefont {E.~R.}\ \bibnamefont {Krausz}}, \ and\
  \bibinfo {author} {\bibfnamefont {D.~J.}\ \bibnamefont {Robbins}},\ }\href
  {\doibase 10.1088/0022-3719/9/13/008} {\bibfield  {journal} {\bibinfo
  {journal} {J. Phys. C: Solid State Phys.}\ }\textbf {\bibinfo {volume} {9}},\
  \bibinfo {pages} {2481} (\bibinfo {year} {1976})}\BibitemShut {NoStop}%
\bibitem [{\citenamefont {Ni}\ \emph {et~al.}(2021)\citenamefont {Ni},
  \citenamefont {Li}, \citenamefont {Amoroso}, \citenamefont {He},
  \citenamefont {Feng}, \citenamefont {Kan}, \citenamefont {Picozzi},\ and\
  \citenamefont {Xiang}}]{ni2021giant}%
  \BibitemOpen
  \bibfield  {author} {\bibinfo {author} {\bibfnamefont {J.~Y.}\ \bibnamefont
  {Ni}}, \bibinfo {author} {\bibfnamefont {X.~Y.}\ \bibnamefont {Li}}, \bibinfo
  {author} {\bibfnamefont {D.}~\bibnamefont {Amoroso}}, \bibinfo {author}
  {\bibfnamefont {X.}~\bibnamefont {He}}, \bibinfo {author} {\bibfnamefont
  {J.~S.}\ \bibnamefont {Feng}}, \bibinfo {author} {\bibfnamefont {E.~J.}\
  \bibnamefont {Kan}}, \bibinfo {author} {\bibfnamefont {S.}~\bibnamefont
  {Picozzi}}, \ and\ \bibinfo {author} {\bibfnamefont {H.~J.}\ \bibnamefont
  {Xiang}},\ }\href {\doibase 10.1103/PhysRevLett.127.247204} {\bibfield
  {journal} {\bibinfo  {journal} {Phys. Rev. Lett.}\ }\textbf {\bibinfo
  {volume} {127}},\ \bibinfo {pages} {247204} (\bibinfo {year}
  {2021})}\BibitemShut {NoStop}%
\bibitem [{\citenamefont {Amoroso}\ \emph {et~al.}(2021)\citenamefont
  {Amoroso}, \citenamefont {Barone},\ and\ \citenamefont
  {Picozzi}}]{amoroso2021tuning}%
  \BibitemOpen
  \bibfield  {author} {\bibinfo {author} {\bibfnamefont {D.}~\bibnamefont
  {Amoroso}}, \bibinfo {author} {\bibfnamefont {P.}~\bibnamefont {Barone}}, \
  and\ \bibinfo {author} {\bibfnamefont {S.}~\bibnamefont {Picozzi}},\ }\href
  {\doibase 10.3390/nano11081873} {\bibfield  {journal} {\bibinfo  {journal}
  {Nanomaterials}\ }\textbf {\bibinfo {volume} {11}},\ \bibinfo {pages} {1873}
  (\bibinfo {year} {2021})}\BibitemShut {NoStop}%
\bibitem [{\citenamefont {Katsumata}\ and\ \citenamefont
  {Yamasaka}(1973)}]{katsumata1973effect}%
  \BibitemOpen
  \bibfield  {author} {\bibinfo {author} {\bibfnamefont {K.}~\bibnamefont
  {Katsumata}}\ and\ \bibinfo {author} {\bibfnamefont {K.}~\bibnamefont
  {Yamasaka}},\ }\href@noop {} {\bibfield  {journal} {\bibinfo  {journal} {J.
  Phys. Soc. Jpn.}\ }\textbf {\bibinfo {volume} {34}},\ \bibinfo {pages} {346}
  (\bibinfo {year} {1973})}\BibitemShut {NoStop}%
\bibitem [{\citenamefont {Billerey}\ \emph
  {et~al.}(1980{\natexlab{a}})\citenamefont {Billerey}, \citenamefont
  {Terrier}, \citenamefont {Pointon},\ and\ \citenamefont
  {Redoules}}]{billerey1980low}%
  \BibitemOpen
  \bibfield  {author} {\bibinfo {author} {\bibfnamefont {D.}~\bibnamefont
  {Billerey}}, \bibinfo {author} {\bibfnamefont {C.}~\bibnamefont {Terrier}},
  \bibinfo {author} {\bibfnamefont {A.~J.}\ \bibnamefont {Pointon}}, \ and\
  \bibinfo {author} {\bibfnamefont {J.~P.}\ \bibnamefont {Redoules}},\
  }\href@noop {} {\bibfield  {journal} {\bibinfo  {journal} {J. Magn. Magn.
  Mater.}\ }\textbf {\bibinfo {volume} {21}},\ \bibinfo {pages} {187} (\bibinfo
  {year} {1980}{\natexlab{a}})}\BibitemShut {NoStop}%
\bibitem [{\citenamefont {Xu}\ \emph {et~al.}(2020)\citenamefont {Xu},
  \citenamefont {Feng}, \citenamefont {Kawamura}, \citenamefont {Yamaji},
  \citenamefont {Nahas}, \citenamefont {Prokhorenko}, \citenamefont {Qi},
  \citenamefont {Xiang},\ and\ \citenamefont {Bellaiche}}]{xu2020}%
  \BibitemOpen
  \bibfield  {author} {\bibinfo {author} {\bibfnamefont {C.}~\bibnamefont
  {Xu}}, \bibinfo {author} {\bibfnamefont {J.}~\bibnamefont {Feng}}, \bibinfo
  {author} {\bibfnamefont {M.}~\bibnamefont {Kawamura}}, \bibinfo {author}
  {\bibfnamefont {Y.}~\bibnamefont {Yamaji}}, \bibinfo {author} {\bibfnamefont
  {Y.}~\bibnamefont {Nahas}}, \bibinfo {author} {\bibfnamefont
  {S.}~\bibnamefont {Prokhorenko}}, \bibinfo {author} {\bibfnamefont
  {Y.}~\bibnamefont {Qi}}, \bibinfo {author} {\bibfnamefont {H.}~\bibnamefont
  {Xiang}}, \ and\ \bibinfo {author} {\bibfnamefont {L.}~\bibnamefont
  {Bellaiche}},\ }\href {\doibase 10.1103/PhysRevLett.124.087205} {\bibfield
  {journal} {\bibinfo  {journal} {Phys. Rev. Lett.}\ }\textbf {\bibinfo
  {volume} {124}},\ \bibinfo {pages} {087205} (\bibinfo {year}
  {2020})}\BibitemShut {NoStop}%
\bibitem [{\citenamefont {R{\'e}gnault}\ \emph {et~al.}(1982)\citenamefont
  {R{\'e}gnault}, \citenamefont {Rossat-Mignod}, \citenamefont {Adam},
  \citenamefont {Billerey},\ and\ \citenamefont
  {Terrier}}]{regnault1982inelastic}%
  \BibitemOpen
  \bibfield  {author} {\bibinfo {author} {\bibfnamefont {L.}~\bibnamefont
  {R{\'e}gnault}}, \bibinfo {author} {\bibfnamefont {J.}~\bibnamefont
  {Rossat-Mignod}}, \bibinfo {author} {\bibfnamefont {A.}~\bibnamefont {Adam}},
  \bibinfo {author} {\bibfnamefont {D.}~\bibnamefont {Billerey}}, \ and\
  \bibinfo {author} {\bibfnamefont {C.}~\bibnamefont {Terrier}},\ }\href@noop
  {} {\bibfield  {journal} {\bibinfo  {journal} {J. physique}\ }\textbf
  {\bibinfo {volume} {43}},\ \bibinfo {pages} {1283} (\bibinfo {year}
  {1982})}\BibitemShut {NoStop}%
\bibitem [{\citenamefont {Rastelli}\ \emph {et~al.}(1979)\citenamefont
  {Rastelli}, \citenamefont {Tassi},\ and\ \citenamefont
  {Reatto}}]{rastelli1979non}%
  \BibitemOpen
  \bibfield  {author} {\bibinfo {author} {\bibfnamefont {E.}~\bibnamefont
  {Rastelli}}, \bibinfo {author} {\bibfnamefont {A.}~\bibnamefont {Tassi}}, \
  and\ \bibinfo {author} {\bibfnamefont {L.}~\bibnamefont {Reatto}},\
  }\href@noop {} {\bibfield  {journal} {\bibinfo  {journal} {Physica B+C}\
  }\textbf {\bibinfo {volume} {97}},\ \bibinfo {pages} {1} (\bibinfo {year}
  {1979})}\BibitemShut {NoStop}%
\bibitem [{\citenamefont {Katsumata}\ and\ \citenamefont
  {Date}(1969)}]{katsumata1969antiferromagnetic}%
  \BibitemOpen
  \bibfield  {author} {\bibinfo {author} {\bibfnamefont {K.}~\bibnamefont
  {Katsumata}}\ and\ \bibinfo {author} {\bibfnamefont {M.}~\bibnamefont
  {Date}},\ }\href {\doibase 10.1143/JPSJ.27.1360} {\bibfield  {journal}
  {\bibinfo  {journal} {J. Phys. Soc. Jpn.}\ }\textbf {\bibinfo {volume}
  {27}},\ \bibinfo {pages} {1360} (\bibinfo {year} {1969})}\BibitemShut
  {NoStop}%
\bibitem [{\citenamefont {Babu}\ \emph {et~al.}(2019)\citenamefont {Babu},
  \citenamefont {Prokeš}, \citenamefont {Huang}, \citenamefont {Radu},\ and\
  \citenamefont {Mishra}}]{babu2019magnetic}%
  \BibitemOpen
  \bibfield  {author} {\bibinfo {author} {\bibfnamefont {S.}~\bibnamefont
  {Babu}}, \bibinfo {author} {\bibfnamefont {K.}~\bibnamefont {Prokeš}},
  \bibinfo {author} {\bibfnamefont {Y.~K.}\ \bibnamefont {Huang}}, \bibinfo
  {author} {\bibfnamefont {F.}~\bibnamefont {Radu}}, \ and\ \bibinfo {author}
  {\bibfnamefont {S.~K.}\ \bibnamefont {Mishra}},\ }\href {\doibase
  10.1063/1.5066625} {\bibfield  {journal} {\bibinfo  {journal} {J. Appl.
  Phys.}\ }\textbf {\bibinfo {volume} {125}},\ \bibinfo {pages} {093902}
  (\bibinfo {year} {2019})}\BibitemShut {NoStop}%
\bibitem [{\citenamefont {Billerey}\ \emph {et~al.}(1977)\citenamefont
  {Billerey}, \citenamefont {Terrier}, \citenamefont {Ciret},\ and\
  \citenamefont {Kleinclauss}}]{billerey1977neutron}%
  \BibitemOpen
  \bibfield  {author} {\bibinfo {author} {\bibfnamefont {D.}~\bibnamefont
  {Billerey}}, \bibinfo {author} {\bibfnamefont {C.}~\bibnamefont {Terrier}},
  \bibinfo {author} {\bibfnamefont {N.}~\bibnamefont {Ciret}}, \ and\ \bibinfo
  {author} {\bibfnamefont {J.}~\bibnamefont {Kleinclauss}},\ }\href {\doibase
  https://doi.org/10.1016/0375-9601(77)90863-5} {\bibfield  {journal} {\bibinfo
   {journal} {Phys. Lett. A}\ }\textbf {\bibinfo {volume} {61}},\ \bibinfo
  {pages} {138} (\bibinfo {year} {1977})}\BibitemShut {NoStop}%
\bibitem [{\citenamefont {Billerey}\ \emph
  {et~al.}(1980{\natexlab{b}})\citenamefont {Billerey}, \citenamefont
  {Terrier}, \citenamefont {Mainard},\ and\ \citenamefont
  {Pointon}}]{billerey1980magnetic}%
  \BibitemOpen
  \bibfield  {author} {\bibinfo {author} {\bibfnamefont {D.}~\bibnamefont
  {Billerey}}, \bibinfo {author} {\bibfnamefont {C.}~\bibnamefont {Terrier}},
  \bibinfo {author} {\bibfnamefont {R.}~\bibnamefont {Mainard}}, \ and\
  \bibinfo {author} {\bibfnamefont {A.~J.}\ \bibnamefont {Pointon}},\ }\href
  {\doibase https://doi.org/10.1016/0375-9601(80)90636-2} {\bibfield  {journal}
  {\bibinfo  {journal} {Phys. Lett. A}\ }\textbf {\bibinfo {volume} {77}},\
  \bibinfo {pages} {59} (\bibinfo {year} {1980}{\natexlab{b}})}\BibitemShut
  {NoStop}%
\bibitem [{\citenamefont {Kuindersma}\ \emph {et~al.}(1981)\citenamefont
  {Kuindersma}, \citenamefont {Sanchez},\ and\ \citenamefont
  {Haas}}]{kuindersma1981magnetic}%
  \BibitemOpen
  \bibfield  {author} {\bibinfo {author} {\bibfnamefont {S.~R.}\ \bibnamefont
  {Kuindersma}}, \bibinfo {author} {\bibfnamefont {J.~P.}\ \bibnamefont
  {Sanchez}}, \ and\ \bibinfo {author} {\bibfnamefont {C.}~\bibnamefont
  {Haas}},\ }\href {\doibase https://doi.org/10.1016/0378-4363(81)90100-5}
  {\bibfield  {journal} {\bibinfo  {journal} {Physica B+C}\ }\textbf {\bibinfo
  {volume} {111}},\ \bibinfo {pages} {231} (\bibinfo {year}
  {1981})}\BibitemShut {NoStop}%
\bibitem [{\citenamefont {Day}\ and\ \citenamefont
  {Ziebeck}(1980)}]{day1980incommensurate}%
  \BibitemOpen
  \bibfield  {author} {\bibinfo {author} {\bibfnamefont {P.}~\bibnamefont
  {Day}}\ and\ \bibinfo {author} {\bibfnamefont {K.~R.~A.}\ \bibnamefont
  {Ziebeck}},\ }\href {\doibase 10.1088/0022-3719/13/21/005} {\bibfield
  {journal} {\bibinfo  {journal} {J. Phys. C: Solid State Phys.}\ }\textbf
  {\bibinfo {volume} {13}},\ \bibinfo {pages} {L523} (\bibinfo {year}
  {1980})}\BibitemShut {NoStop}%
\bibitem [{\citenamefont {Adam}\ \emph {et~al.}(1980)\citenamefont {Adam},
  \citenamefont {Billerey}, \citenamefont {Terrier}, \citenamefont {Mainard},
  \citenamefont {Regnault}, \citenamefont {Rossat-Mignod},\ and\ \citenamefont
  {M\'eriel}}]{adam1980neutron}%
  \BibitemOpen
  \bibfield  {author} {\bibinfo {author} {\bibfnamefont {A.}~\bibnamefont
  {Adam}}, \bibinfo {author} {\bibfnamefont {D.}~\bibnamefont {Billerey}},
  \bibinfo {author} {\bibfnamefont {C.}~\bibnamefont {Terrier}}, \bibinfo
  {author} {\bibfnamefont {R.}~\bibnamefont {Mainard}}, \bibinfo {author}
  {\bibfnamefont {L.~P.}\ \bibnamefont {Regnault}}, \bibinfo {author}
  {\bibfnamefont {J.}~\bibnamefont {Rossat-Mignod}}, \ and\ \bibinfo {author}
  {\bibfnamefont {P.}~\bibnamefont {M\'eriel}},\ }\href {\doibase
  https://doi.org/10.1016/0038-1098(80)90757-7} {\bibfield  {journal} {\bibinfo
   {journal} {Solid State Commun.}\ }\textbf {\bibinfo {volume} {35}},\
  \bibinfo {pages} {1} (\bibinfo {year} {1980})}\BibitemShut {NoStop}%
\bibitem [{\citenamefont {Day}\ \emph {et~al.}(1984)\citenamefont {Day},
  \citenamefont {Moore}, \citenamefont {Wood}, \citenamefont {Paul},
  \citenamefont {Ziebeck}, \citenamefont {Regnault},\ and\ \citenamefont
  {Rossat-Mignod}}]{day1984inelastic}%
  \BibitemOpen
  \bibfield  {author} {\bibinfo {author} {\bibfnamefont {P.}~\bibnamefont
  {Day}}, \bibinfo {author} {\bibfnamefont {M.~W.}\ \bibnamefont {Moore}},
  \bibinfo {author} {\bibfnamefont {T.~E.}\ \bibnamefont {Wood}}, \bibinfo
  {author} {\bibfnamefont {D.~M.}\ \bibnamefont {Paul}}, \bibinfo {author}
  {\bibfnamefont {K.~R.}\ \bibnamefont {Ziebeck}}, \bibinfo {author}
  {\bibfnamefont {L.~P.}\ \bibnamefont {Regnault}}, \ and\ \bibinfo {author}
  {\bibfnamefont {J.}~\bibnamefont {Rossat-Mignod}},\ }\href@noop {} {\bibfield
   {journal} {\bibinfo  {journal} {Solid State Commun.}\ }\textbf {\bibinfo
  {volume} {51}},\ \bibinfo {pages} {627} (\bibinfo {year} {1984})}\BibitemShut
  {NoStop}%
\bibitem [{\citenamefont {Riedl}\ \emph {et~al.}(2021)\citenamefont {Riedl},
  \citenamefont {Gati}, \citenamefont {Zielke}, \citenamefont {Hartmann},
  \citenamefont {Vyaselev}, \citenamefont {Kushch}, \citenamefont {Jeschke},
  \citenamefont {Lang}, \citenamefont {Valent\'{\i}}, \citenamefont
  {Kartsovnik},\ and\ \citenamefont {Winter}}]{riedl2021spin}%
  \BibitemOpen
  \bibfield  {author} {\bibinfo {author} {\bibfnamefont {K.}~\bibnamefont
  {Riedl}}, \bibinfo {author} {\bibfnamefont {E.}~\bibnamefont {Gati}},
  \bibinfo {author} {\bibfnamefont {D.}~\bibnamefont {Zielke}}, \bibinfo
  {author} {\bibfnamefont {S.}~\bibnamefont {Hartmann}}, \bibinfo {author}
  {\bibfnamefont {O.~M.}\ \bibnamefont {Vyaselev}}, \bibinfo {author}
  {\bibfnamefont {N.~D.}\ \bibnamefont {Kushch}}, \bibinfo {author}
  {\bibfnamefont {H.~O.}\ \bibnamefont {Jeschke}}, \bibinfo {author}
  {\bibfnamefont {M.}~\bibnamefont {Lang}}, \bibinfo {author} {\bibfnamefont
  {R.}~\bibnamefont {Valent\'{\i}}}, \bibinfo {author} {\bibfnamefont {M.~V.}\
  \bibnamefont {Kartsovnik}}, \ and\ \bibinfo {author} {\bibfnamefont {S.~M.}\
  \bibnamefont {Winter}},\ }\href {\doibase 10.1103/PhysRevLett.127.147204}
  {\bibfield  {journal} {\bibinfo  {journal} {Phys. Rev. Lett.}\ }\textbf
  {\bibinfo {volume} {127}},\ \bibinfo {pages} {147204} (\bibinfo {year}
  {2021})}\BibitemShut {NoStop}%
\bibitem [{\citenamefont {Niel}\ \emph {et~al.}(1977)\citenamefont {Niel},
  \citenamefont {Cros}, \citenamefont {Le~Flem}, \citenamefont {Pouchard},\
  and\ \citenamefont {Hagenmuller}}]{niel1977magnetic}%
  \BibitemOpen
  \bibfield  {author} {\bibinfo {author} {\bibfnamefont {M.}~\bibnamefont
  {Niel}}, \bibinfo {author} {\bibfnamefont {C.}~\bibnamefont {Cros}}, \bibinfo
  {author} {\bibfnamefont {G.}~\bibnamefont {Le~Flem}}, \bibinfo {author}
  {\bibfnamefont {M.}~\bibnamefont {Pouchard}}, \ and\ \bibinfo {author}
  {\bibfnamefont {P.}~\bibnamefont {Hagenmuller}},\ }\href@noop {} {\bibfield
  {journal} {\bibinfo  {journal} {Physica B+C}\ }\textbf {\bibinfo {volume}
  {86}},\ \bibinfo {pages} {702} (\bibinfo {year} {1977})}\BibitemShut
  {NoStop}%
\bibitem [{\citenamefont {Kadowaki}\ \emph {et~al.}(1987)\citenamefont
  {Kadowaki}, \citenamefont {Ubukoshi}, \citenamefont {Hirakawa}, \citenamefont
  {Mart{\'\i}nez},\ and\ \citenamefont {Shirane}}]{kadowaki1987experimental}%
  \BibitemOpen
  \bibfield  {author} {\bibinfo {author} {\bibfnamefont {H.}~\bibnamefont
  {Kadowaki}}, \bibinfo {author} {\bibfnamefont {K.}~\bibnamefont {Ubukoshi}},
  \bibinfo {author} {\bibfnamefont {K.}~\bibnamefont {Hirakawa}}, \bibinfo
  {author} {\bibfnamefont {J.~L.}\ \bibnamefont {Mart{\'\i}nez}}, \ and\
  \bibinfo {author} {\bibfnamefont {G.}~\bibnamefont {Shirane}},\ }\href@noop
  {} {\bibfield  {journal} {\bibinfo  {journal} {J. Phys. Soc. Jpn.}\ }\textbf
  {\bibinfo {volume} {56}},\ \bibinfo {pages} {4027} (\bibinfo {year}
  {1987})}\BibitemShut {NoStop}%
\bibitem [{\citenamefont {Kadowaki}\ \emph {et~al.}(1985)\citenamefont
  {Kadowaki}, \citenamefont {Ubukoshi},\ and\ \citenamefont
  {Hirakawa}}]{kadowaki1985neutron}%
  \BibitemOpen
  \bibfield  {author} {\bibinfo {author} {\bibfnamefont {H.}~\bibnamefont
  {Kadowaki}}, \bibinfo {author} {\bibfnamefont {K.}~\bibnamefont {Ubukoshi}},
  \ and\ \bibinfo {author} {\bibfnamefont {K.}~\bibnamefont {Hirakawa}},\
  }\href@noop {} {\bibfield  {journal} {\bibinfo  {journal} {J. Phys. Soc.
  Jpn.}\ }\textbf {\bibinfo {volume} {54}},\ \bibinfo {pages} {363} (\bibinfo
  {year} {1985})}\BibitemShut {NoStop}%
\bibitem [{\citenamefont {Hirakawa}\ \emph {et~al.}(1983)\citenamefont
  {Hirakawa}, \citenamefont {Kadowaki},\ and\ \citenamefont
  {Ubukoshi}}]{hirakawa1983study}%
  \BibitemOpen
  \bibfield  {author} {\bibinfo {author} {\bibfnamefont {K.}~\bibnamefont
  {Hirakawa}}, \bibinfo {author} {\bibfnamefont {H.}~\bibnamefont {Kadowaki}},
  \ and\ \bibinfo {author} {\bibfnamefont {K.}~\bibnamefont {Ubukoshi}},\
  }\href@noop {} {\bibfield  {journal} {\bibinfo  {journal} {J. Phys. Soc.
  Jpn.}\ }\textbf {\bibinfo {volume} {52}},\ \bibinfo {pages} {1814} (\bibinfo
  {year} {1983})}\BibitemShut {NoStop}%
\bibitem [{\citenamefont {Nishi}\ \emph {et~al.}(1984)\citenamefont {Nishi},
  \citenamefont {Ito}, \citenamefont {Kadowaki},\ and\ \citenamefont
  {Hirakawa}}]{nishi1984neutron}%
  \BibitemOpen
  \bibfield  {author} {\bibinfo {author} {\bibfnamefont {M.}~\bibnamefont
  {Nishi}}, \bibinfo {author} {\bibfnamefont {Y.}~\bibnamefont {Ito}}, \bibinfo
  {author} {\bibfnamefont {H.}~\bibnamefont {Kadowaki}}, \ and\ \bibinfo
  {author} {\bibfnamefont {K.}~\bibnamefont {Hirakawa}},\ }\href@noop {}
  {\bibfield  {journal} {\bibinfo  {journal} {J. Phys. Soc. Jpn.}\ }\textbf
  {\bibinfo {volume} {53}},\ \bibinfo {pages} {1214} (\bibinfo {year}
  {1984})}\BibitemShut {NoStop}%
\bibitem [{\citenamefont {Kuindersma}\ \emph {et~al.}(1979)\citenamefont
  {Kuindersma}, \citenamefont {Haas}, \citenamefont {Sanchez},\ and\
  \citenamefont {Al}}]{kuindersma1979magnetic}%
  \BibitemOpen
  \bibfield  {author} {\bibinfo {author} {\bibfnamefont {S.}~\bibnamefont
  {Kuindersma}}, \bibinfo {author} {\bibfnamefont {C.}~\bibnamefont {Haas}},
  \bibinfo {author} {\bibfnamefont {J.}~\bibnamefont {Sanchez}}, \ and\
  \bibinfo {author} {\bibfnamefont {R.}~\bibnamefont {Al}},\ }\href@noop {}
  {\bibfield  {journal} {\bibinfo  {journal} {Solid State Commun.}\ }\textbf
  {\bibinfo {volume} {30}},\ \bibinfo {pages} {403} (\bibinfo {year}
  {1979})}\BibitemShut {NoStop}%
\bibitem [{ost(1958)}]{osti_4344362}%
  \BibitemOpen
  \href {\doibase 10.2172/4344362} {\emph {\bibinfo {title} {Physics division
  semiannual progress report for period ending September 10, 1957}}},\ \bibinfo
  {type} {Tech. Rep.}\ (\bibinfo  {institution} {Oak Ridge National Lab.,
  Tenn.},\ \bibinfo {year} {1958})\BibitemShut {NoStop}%
\bibitem [{\citenamefont {Sato}\ \emph {et~al.}(1995)\citenamefont {Sato},
  \citenamefont {Kadowaki},\ and\ \citenamefont {Iio}}]{sato1995successive}%
  \BibitemOpen
  \bibfield  {author} {\bibinfo {author} {\bibfnamefont {T.}~\bibnamefont
  {Sato}}, \bibinfo {author} {\bibfnamefont {H.}~\bibnamefont {Kadowaki}}, \
  and\ \bibinfo {author} {\bibfnamefont {K.}~\bibnamefont {Iio}},\ }\href@noop
  {} {\bibfield  {journal} {\bibinfo  {journal} {Physica B: Condens. Matter}\
  }\textbf {\bibinfo {volume} {213}},\ \bibinfo {pages} {224} (\bibinfo {year}
  {1995})}\BibitemShut {NoStop}%
\bibitem [{\citenamefont {Iio}\ \emph {et~al.}(1990)\citenamefont {Iio},
  \citenamefont {Masuda}, \citenamefont {Tanaka},\ and\ \citenamefont
  {Nagata}}]{iio1990successive}%
  \BibitemOpen
  \bibfield  {author} {\bibinfo {author} {\bibfnamefont {K.}~\bibnamefont
  {Iio}}, \bibinfo {author} {\bibfnamefont {H.}~\bibnamefont {Masuda}},
  \bibinfo {author} {\bibfnamefont {H.}~\bibnamefont {Tanaka}}, \ and\ \bibinfo
  {author} {\bibfnamefont {K.}~\bibnamefont {Nagata}},\ }\href@noop {}
  {\bibfield  {journal} {\bibinfo  {journal} {J. Magn. Magn. Mater.}\ }\textbf
  {\bibinfo {volume} {90}},\ \bibinfo {pages} {265} (\bibinfo {year}
  {1990})}\BibitemShut {NoStop}%
\bibitem [{\citenamefont {Sato}\ and\ \citenamefont
  {Kadowaki}(1993)}]{sato1993neutron}%
  \BibitemOpen
  \bibfield  {author} {\bibinfo {author} {\bibfnamefont {T.}~\bibnamefont
  {Sato}}\ and\ \bibinfo {author} {\bibfnamefont {H.}~\bibnamefont
  {Kadowaki}},\ }\href@noop {} {\emph {\bibinfo {title} {Neutron scattering
  study of MnX$_2$ (X= Br, I)}}},\ \bibinfo {type} {Tech. Rep.}\ (\bibinfo
  {year} {1993})\BibitemShut {NoStop}%
\bibitem [{\citenamefont {Cable}\ \emph {et~al.}(1962)\citenamefont {Cable},
  \citenamefont {Wilkinson}, \citenamefont {Wollan},\ and\ \citenamefont
  {Koehler}}]{cable1962neutron}%
  \BibitemOpen
  \bibfield  {author} {\bibinfo {author} {\bibfnamefont {J.}~\bibnamefont
  {Cable}}, \bibinfo {author} {\bibfnamefont {M.}~\bibnamefont {Wilkinson}},
  \bibinfo {author} {\bibfnamefont {E.}~\bibnamefont {Wollan}}, \ and\ \bibinfo
  {author} {\bibfnamefont {W.}~\bibnamefont {Koehler}},\ }\href@noop {}
  {\bibfield  {journal} {\bibinfo  {journal} {Phys. Rev.}\ }\textbf {\bibinfo
  {volume} {125}},\ \bibinfo {pages} {1860} (\bibinfo {year}
  {1962})}\BibitemShut {NoStop}%
\bibitem [{\citenamefont {Utesov}\ and\ \citenamefont
  {Syromyatnikov}(2017)}]{utesov2017cascades}%
  \BibitemOpen
  \bibfield  {author} {\bibinfo {author} {\bibfnamefont {O.~I.}\ \bibnamefont
  {Utesov}}\ and\ \bibinfo {author} {\bibfnamefont {A.~V.}\ \bibnamefont
  {Syromyatnikov}},\ }\href {\doibase 10.1103/PhysRevB.95.214420} {\bibfield
  {journal} {\bibinfo  {journal} {Phys. Rev. B}\ }\textbf {\bibinfo {volume}
  {95}},\ \bibinfo {pages} {214420} (\bibinfo {year} {2017})}\BibitemShut
  {NoStop}%
\bibitem [{\citenamefont {Kresse}\ and\ \citenamefont
  {Furthm\"uller}(1996)}]{kresse_paw}%
  \BibitemOpen
  \bibfield  {author} {\bibinfo {author} {\bibfnamefont {G.}~\bibnamefont
  {Kresse}}\ and\ \bibinfo {author} {\bibfnamefont {J.}~\bibnamefont
  {Furthm\"uller}},\ }\href {\doibase 10.1103/PhysRevB.54.11169} {\bibfield
  {journal} {\bibinfo  {journal} {Phys. Rev. B}\ }\textbf {\bibinfo {volume}
  {54}},\ \bibinfo {pages} {11169} (\bibinfo {year} {1996})}\BibitemShut
  {NoStop}%
\bibitem [{\citenamefont {McGuire}(2017)}]{mcguire_rev}%
  \BibitemOpen
  \bibfield  {author} {\bibinfo {author} {\bibfnamefont {M.~A.}\ \bibnamefont
  {McGuire}},\ }\href {\doibase 10.3390/cryst7050121} {\bibfield  {journal}
  {\bibinfo  {journal} {Crystals}\ }\textbf {\bibinfo {volume} {7}},\ \bibinfo
  {pages} {121} (\bibinfo {year} {2017})}\BibitemShut {NoStop}%
\bibitem [{\citenamefont {Jiang}\ \emph {et~al.}(2012)\citenamefont {Jiang},
  \citenamefont {G\'omez-Abal}, \citenamefont {Li}, \citenamefont
  {Meisenbichler}, \citenamefont {Ambrosch-Draxl},\ and\ \citenamefont
  {Scheffler}}]{fhigap}%
  \BibitemOpen
  \bibfield  {author} {\bibinfo {author} {\bibfnamefont {H.}~\bibnamefont
  {Jiang}}, \bibinfo {author} {\bibfnamefont {R.~I.}\ \bibnamefont
  {G\'omez-Abal}}, \bibinfo {author} {\bibfnamefont {X.}~\bibnamefont {Li}},
  \bibinfo {author} {\bibfnamefont {C.}~\bibnamefont {Meisenbichler}}, \bibinfo
  {author} {\bibfnamefont {C.}~\bibnamefont {Ambrosch-Draxl}}, \ and\ \bibinfo
  {author} {\bibfnamefont {M.}~\bibnamefont {Scheffler}},\ }\href@noop {}
  {\bibfield  {journal} {\bibinfo  {journal} {Computer Phys. Commun.}\ }\textbf
  {\bibinfo {volume} {184}},\ \bibinfo {pages} {348} (\bibinfo {year}
  {2012})}\BibitemShut {NoStop}%
\bibitem [{\citenamefont {Yekta}\ \emph {et~al.}(2021)\citenamefont {Yekta},
  \citenamefont {Hadipour}, \citenamefont {\ifmmode \mbox{\c{S}}\else
  \c{S}\fi{}a\ifmmode \mbox{\c{s}}\else \c{s}\fi{}\ifmmode \imath \else \i
  \fi{}o\ifmmode~\breve{g}\else \u{g}\fi{}lu}, \citenamefont {Friedrich},
  \citenamefont {Jafari}, \citenamefont {Bl\"ugel},\ and\ \citenamefont
  {Mertig}}]{yekta2021strength}%
  \BibitemOpen
  \bibfield  {author} {\bibinfo {author} {\bibfnamefont {Y.}~\bibnamefont
  {Yekta}}, \bibinfo {author} {\bibfnamefont {H.}~\bibnamefont {Hadipour}},
  \bibinfo {author} {\bibfnamefont {E.}~\bibnamefont {\ifmmode
  \mbox{\c{S}}\else \c{S}\fi{}a\ifmmode \mbox{\c{s}}\else \c{s}\fi{}\ifmmode
  \imath \else \i \fi{}o\ifmmode~\breve{g}\else \u{g}\fi{}lu}}, \bibinfo
  {author} {\bibfnamefont {C.}~\bibnamefont {Friedrich}}, \bibinfo {author}
  {\bibfnamefont {S.~A.}\ \bibnamefont {Jafari}}, \bibinfo {author}
  {\bibfnamefont {S.}~\bibnamefont {Bl\"ugel}}, \ and\ \bibinfo {author}
  {\bibfnamefont {I.}~\bibnamefont {Mertig}},\ }\href {\doibase
  10.1103/PhysRevMaterials.5.034001} {\bibfield  {journal} {\bibinfo  {journal}
  {Phys. Rev. Materials}\ }\textbf {\bibinfo {volume} {5}},\ \bibinfo {pages}
  {034001} (\bibinfo {year} {2021})}\BibitemShut {NoStop}%
\bibitem [{\citenamefont {Liechtenstein}\ \emph {et~al.}(1995)\citenamefont
  {Liechtenstein}, \citenamefont {Anisimov},\ and\ \citenamefont
  {Zaanen}}]{liechtenstein1995density}%
  \BibitemOpen
  \bibfield  {author} {\bibinfo {author} {\bibfnamefont {A.~I.}\ \bibnamefont
  {Liechtenstein}}, \bibinfo {author} {\bibfnamefont {V.~I.}\ \bibnamefont
  {Anisimov}}, \ and\ \bibinfo {author} {\bibfnamefont {J.}~\bibnamefont
  {Zaanen}},\ }\href {\doibase 10.1103/PhysRevB.52.R5467} {\bibfield  {journal}
  {\bibinfo  {journal} {Phys. Rev. B}\ }\textbf {\bibinfo {volume} {52}},\
  \bibinfo {pages} {R5467} (\bibinfo {year} {1995})}\BibitemShut {NoStop}%
\bibitem [{\citenamefont {Liu}\ and\ \citenamefont
  {Khaliullin}(2018)}]{liu2018pseudospin}%
  \BibitemOpen
  \bibfield  {author} {\bibinfo {author} {\bibfnamefont {H.}~\bibnamefont
  {Liu}}\ and\ \bibinfo {author} {\bibfnamefont {G.}~\bibnamefont
  {Khaliullin}},\ }\href@noop {} {\bibfield  {journal} {\bibinfo  {journal}
  {Phys. Rev. B}\ }\textbf {\bibinfo {volume} {97}},\ \bibinfo {pages} {014407}
  (\bibinfo {year} {2018})}\BibitemShut {NoStop}%
\bibitem [{\citenamefont {Winter}\ \emph
  {et~al.}(2017{\natexlab{c}})\citenamefont {Winter}, \citenamefont {Riedl},\
  and\ \citenamefont {Valent\'{\i}}}]{winter2017importance}%
  \BibitemOpen
  \bibfield  {author} {\bibinfo {author} {\bibfnamefont {S.~M.}\ \bibnamefont
  {Winter}}, \bibinfo {author} {\bibfnamefont {K.}~\bibnamefont {Riedl}}, \
  and\ \bibinfo {author} {\bibfnamefont {R.}~\bibnamefont {Valent\'{\i}}},\
  }\href {\doibase 10.1103/PhysRevB.95.060404} {\bibfield  {journal} {\bibinfo
  {journal} {Phys. Rev. B}\ }\textbf {\bibinfo {volume} {95}},\ \bibinfo
  {pages} {060404} (\bibinfo {year} {2017}{\natexlab{c}})}\BibitemShut
  {NoStop}%
\bibitem [{\citenamefont {Rau}\ \emph {et~al.}(2014)\citenamefont {Rau},
  \citenamefont {Lee},\ and\ \citenamefont {Kee}}]{rau2014generic}%
  \BibitemOpen
  \bibfield  {author} {\bibinfo {author} {\bibfnamefont {J.~G.}\ \bibnamefont
  {Rau}}, \bibinfo {author} {\bibfnamefont {E.~K.-H.}\ \bibnamefont {Lee}}, \
  and\ \bibinfo {author} {\bibfnamefont {H.-Y.}\ \bibnamefont {Kee}},\ }\href
  {\doibase 10.1103/PhysRevLett.112.077204} {\bibfield  {journal} {\bibinfo
  {journal} {Phys. Rev. Lett.}\ }\textbf {\bibinfo {volume} {112}},\ \bibinfo
  {pages} {077204} (\bibinfo {year} {2014})}\BibitemShut {NoStop}%
\bibitem [{\citenamefont {Nguyen}\ \emph {et~al.}(2021)\citenamefont {Nguyen},
  \citenamefont {Yamauchi}, \citenamefont {Oguchi}, \citenamefont {Amoroso},\
  and\ \citenamefont {Picozzi}}]{thao2021}%
  \BibitemOpen
  \bibfield  {author} {\bibinfo {author} {\bibfnamefont {T.~P.~T.}\
  \bibnamefont {Nguyen}}, \bibinfo {author} {\bibfnamefont {K.}~\bibnamefont
  {Yamauchi}}, \bibinfo {author} {\bibfnamefont {T.}~\bibnamefont {Oguchi}},
  \bibinfo {author} {\bibfnamefont {D.}~\bibnamefont {Amoroso}}, \ and\
  \bibinfo {author} {\bibfnamefont {S.}~\bibnamefont {Picozzi}},\ }\href
  {\doibase 10.1103/PhysRevB.104.014414} {\bibfield  {journal} {\bibinfo
  {journal} {Phys. Rev. B}\ }\textbf {\bibinfo {volume} {104}},\ \bibinfo
  {pages} {014414} (\bibinfo {year} {2021})}\BibitemShut {NoStop}%
\end{thebibliography}%
\end{document}